\documentclass{emulateapj}

\shorttitle{Sensitivity of $p$ Process Nucleosynthesis
to Nuclear Reaction Rates in a 25 Solar Mass Supernova Model}
\shortauthors{W. Rapp et al.}

\begin{document}

\title{Sensitivity of $p$ Process Nucleosynthesis to Nuclear Reaction Rates in a 25 Solar Mass Supernova Model}

\author{W. Rapp, J. G\"orres, and M. Wiescher}

\affil{University of Notre Dame,
                 Department of Physics \&
                 Joint Institute of Nuclear Astrophysics,
                 225 Nieuwland Science Hall, Notre Dame,
                 IN 46556, USA}


\author{H. Schatz}

\affil{Department of Physics and Astronomy, National
                Superconducting Cyclotron Laboratory \& \\ Joint Institute
                of Nuclear Astrophysics, Michigan State University,
                East Lansing, MI 48824, USA}

\author{F. K\"appeler}

\affil{Forschungszentrum Karlsruhe,
                Institut f\"ur Kernphysik, P.O. Box 3640,
                76021 Karlsruhe, Germany}

\begin{abstract}

The astrophysical {\it p} process, which is responsible for the
origin of the proton rich stable nuclei heavier than iron, was
investigated using a full nuclear reaction network for a type II
supernova explosion when the shock front passes through the O/Ne
layer. Calculations were performed with a multi-layer model
adopting the seed of a pre-explosion evolution of a 25 solar mass
star. The reaction flux was calculated to determine the main
reaction path and branching points responsible for synthesizing
the proton rich nuclei. In order to investigate the impact of
nuclear reaction rates on the predicted {\it p}-process
abundances, extensive simulations with different sets of
collectively and individually modified neutron-, proton-,
$\alpha$-capture and photodisintegration rates have been
performed. These results are not only relevant to explore the
nuclear physics related uncertainties in {\it p}-process
calculations but are also important for identifying the strategy
and planning of future experiments.

\end{abstract}


\section{Introduction}

The majority of the observed heavy nuclei above iron have been
produced by neutron induced nucleosynthesis processes such as the
slow neutron capture process ({\it s} process) in low mass AGB
stars and massive red giant stars and the rapid neutron capture
process ({\it r} process) in the supernova shock front \citep{KBW89}.

In nature, 35
nuclei can be found on the neutron deficient side of the valley of
stability ranging from $^{74}$Se to $^{196}$Hg, which are shielded
against production by neutron capture processes. For the
nucleosynthesis of these isotopes the {\it p} process has been
proposed as the most likely scenario. The nature of the {\it p}
process is still under debate and it is possible that it
represents different independent nucleosynthesis scenarios. 
The contribution from charged particle induced nuclear reactions
are most likely negligible because the high Coulomb barrier
reduces the associated reaction rates significantly, though 
the possibility of 
some contribution to the lightest p-nuclei cannot be entirely 
excluded \citep{Sch98,JoM04}.
In its
current interpretation the {\it p} process is described as
$\gamma$ induced photodisintegration of stable nuclei in the shock
front of type II supernovae \citep{WoH78,RPA90} or alternatively
in the deflagration flame of a type I supernova detonation
\citep{HMW91}. 
More recently pre-explosive sites for the
$p$-process have been suggested in the O-Ne burning zone of
massive stars. 
This feature emerges in one dimensional stellar
evolution models \citep{RauHH02} and also in two
dimensional simulations of convective oxygen rich burning zones.
However, in the two dimensional study
the calculation of the associated nucleosynthesis pattern has been
limited to a one dimensional model
adaptation and does not yet include quantitative abundance
predictions \citep{BaA01}.
Supercritical accretion disks associated with jets
in supernovae have also been proposed as a possible scenario \citep{FHK03}. 

\begin{figure}
epsscale{.80} \plotone{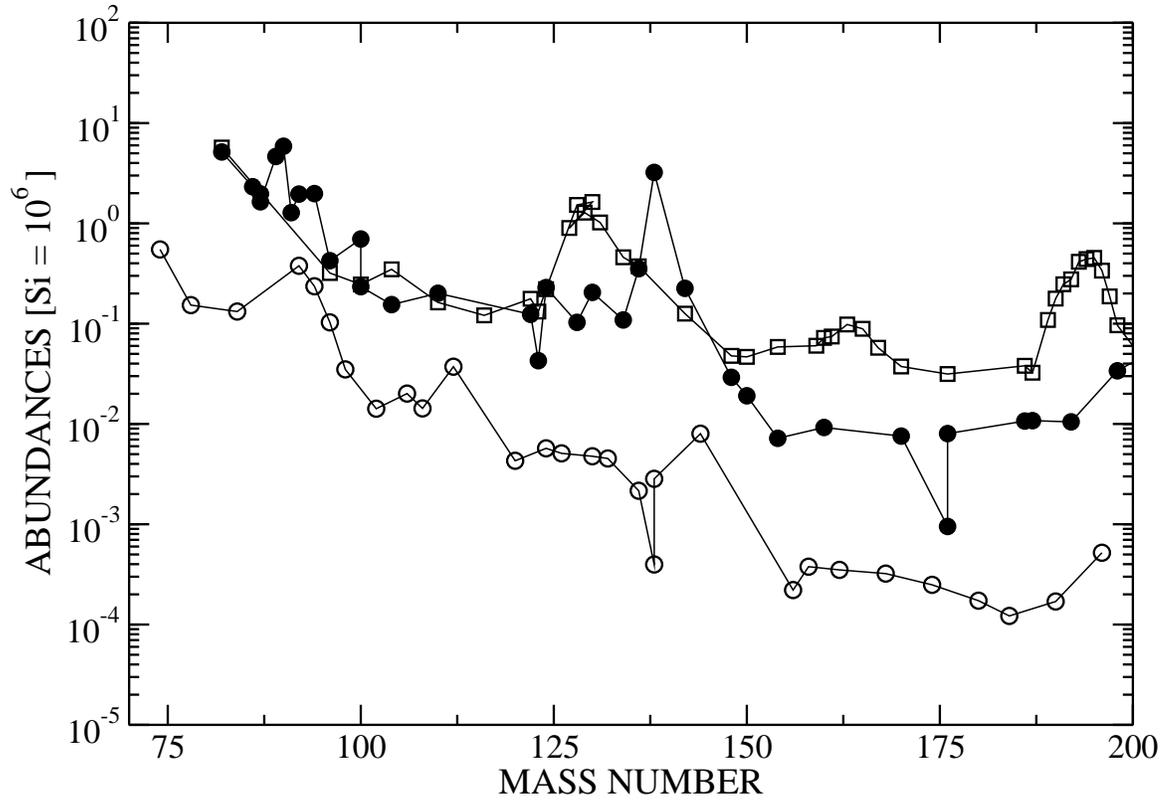}
\caption{Abundance distribution of those heavy isotopes, which can be entirely
ascribed to $s$ process (full circles), $r$ process (open squares), and {\it p}
process (open circles) nucleosynthesis.\label{rapp_fig1}}
\end{figure}

The bulk of
{\it p} isotopes represents a small fraction of the total abundance and is
presently based on the analysis of meteorite data \citep{AnG89}.
This is illustrated in Fig. \ref{rapp_fig1}, where the abundance distributions
are compared for all those heavy
isotopes, which can be entirely ascribed to a specific production
process. In most cases the predicted {\it p}-process abundances agree
within a factor of three with the observed values \citep{Lam92,Mey94,RAH95}.
However, there still are significant discrepancies for
the light {\it p} nuclei
$^{92,94}$Mo and $^{96,98}$Ru, which are largely overabundant
compared to the model predictions,
as well as
for the Dy and Gd p-isotopes in mass region $A=150-160$.
Such discrepancies are reviewed in more detail in \citet{ArG03},
who give a comprehensive review of the
present interpretation and questions related to {\it p}-process
nucleosynthesis. While \citet{ArG03}
discuss the global aspects of {\it p}-process
nucleosynthesis, the present work concentrates on the
nuclear physics related uncertainties in {\it p}-process model
predictions and in particular on the identification of critical
{\it p}-process reaction rates, which are most important for further
experimental studies.

The presently favored scenario for the {\it p} process are type II
supernova (SN) explosions \citep{ArG03,HIS04}. The emerging
shock front causes a rapid increase in temperature and density in
the different layers of the pre-supernova star. As displayed in
Fig. \ref{rapp_fig2} peak temperatures between 1.7$<$T$_9<$3.3 are reached
in the Ne/O layer of the presupernova envelope \citep{RAH95}.
The associated intense photon flux induces a range of
photo-disintegration processes shifting the existing distribution of seed
abundances to the proton rich side of the valley of stability by
($\gamma$,n)-reactions. When this process becomes less
efficient because the neutron binding energy increases with neutron
deficiency, the reaction flux is maintained by the ($\gamma$,p) and
($\gamma$,$\alpha$) channels. After the shock front has passed the Ne/O
layers, temperature and density drop exponentially and the
unstable proton rich nuclei decay back to the valley of stability.
In chapter 3 the {\it p}-process reaction flux will be discussed
more quantitatively in terms of our present model predictions.

The description of the entire synthesis process for the {\it p}
nuclei requires a comprehensive reaction network involving far
more than ten thousand reactions. With very few exceptions the
astrophysical reaction rates have been calculated by means of the
statistical Hauser-Feshbach (HF) model, which can only be applied
for nuclei with high level densities. The statistical model
entries depend on free parameters for the particle potentials and
level densities, as well as for the $\gamma$ widths. These
parameters have to be determined, tested, and improved by
laboratory measurements. The HF predictions do also depend
critically on the masses of the associated nuclei, which are
rather well known in the case of the {\it p} process
\citep{AWT03}. A detailed description of HF techniques for
determining nuclear reaction rates are given in the literature
\citep{RaT00,RaT01,RaT04}. The presently available HF models
describe the astrophysical reaction rates typically only within a
factor of two. The specific limitations with respect to {\it
p}-process analyses have been discussed by \cite{ArG03}.
The
 observed discrepancies between experimental data and theoretical
 HF predictions seem to be largest in the case of $\alpha$ capture
 reactions which has been interpreted as the result of the use of
 insufficient $\alpha$ potential models. We therefore will give
 special consideration to the impact of uncertainties in
 ($\gamma,\alpha$) reactions on $p$-process reaction flow and
 $p$-nuclei abundance predictions.

\section{Experimental Data}

Only a few of the required reaction rates have been investigated
experimentally. While in most cases experimental and theoretical rates
agree within the uncertainties due to the model parameters
\citep{ArG03}, significant discrepancies have been observed in some
interesting cases.

Neutron capture measurements for {\it p}-process studies are available for
part of the $p$ nuclei \cite{BBK00}, whereas methods for the determination
of photodisintegration cross sections and reaction rates have been developed
only recently 
either using inverse Compton scattered laser photons \citep{UYA01} or 
electron bremsstrahlung \citep{VMB01}. The latter
method has been successfully applied to
measure ($\gamma$,n) photodisintegration reactions for
the {\it p} process including the {\it p} nuclei $^{190,192}$Pt
\citep{VMB01}, $^{196,198}$Hg, and the proton magic nucleus $^{204}$Pb
\citep{SVG04}. Good agreement with the HF predictions has been found
in all these cases. Comparable data to test the reliability of
($\gamma$,n) photodisintegration processes in the mass range below
$Z=72$ are not yet available, e.g. for nuclei near the $N=$50 and $N=$82 closed
neutron shells.

Considerable effort has been spent on determining the rates for
($\gamma$,p) and ($\gamma,\alpha$) reactions via measurements of
the inverse capture reactions. The approach has emerged as one of
the major tools for testing the reliability of the HF predictions.
The (p,$\gamma$) rate for the {\it p} isotope $^{96}$Ru is on
average a factor two below the HF results \citep{BSK98}. On the
other hand, experimental studies for the {\it p} nucleus
$^{102}$Pd indicate that the reaction rate for
$^{102}$Pd(p,$\gamma$)$^{103}$Ag is significantly higher than HF
predictions \citep{OMB02}. Recent {\it p}-nuclei related proton
capture studies in the lower mass range such as
$^{74,76}$Se(p,$\gamma$)$^{75,77}$Br \citep{GFS03}, or
$^{92,94}$Mo(p,$\gamma$)$^{93,95}$Tc \citep{SaK97} compare rather
well with HF predictions. Also recent proton capture reaction
studies on neutron magic nuclei with $N=$50 such as
$^{88}$Sr(p,$\gamma$)$^{89}$Y \citep{GDK03} and
$^{89}$Y(p,$\gamma$)$^{90}$Zr \citep{TKS04} seem to agree well
with HF calculations within the parameter space of the model. In
the range $Z\ge$50 experimental proton capture rates for {\it p}
nuclei are not available except for
$^{112}$Sn(p,$\gamma$)$^{113}$Sb \citep{CMB99}, which seems to be
in excellent agreement with HF predictions, and for
$^{116}$Sn(p,$\gamma$)$^{117}$Sb, which showed again significantly
higher experimental values than predicted \citep{OMB02,Rap04}.

Compared to neutron and proton associated capture and
photodisintegration processes, larger deviations from HF predictions
have been reported for $\alpha$ capture and
photodisintegration reactions into the alpha channel. The proton
and neutron binding energies for nuclei along the {\it p}-process path
are typically very large, which warrants the high level density
conditions required for applying the HF model.
For ($\gamma,\alpha$) reactions along the {\it p}-process path,
however, the $\alpha$ binding energy is often low and many
nuclei are even $\alpha$-unbound, spontaneous
$\alpha$-decay being only suppressed by the respective Coulomb
barriers. In these cases, where deviations from HF
predictions might be anticipated, the number of
experimental studies is small and limited to
$\alpha$ capture and scattering reactions on $N=$82 and $N=$50 closed
neutron shell nuclei along the {\it p}-process path. The measurement of
the $^{144}$Sm($\alpha,\gamma$)$^{158}$Gd reaction (Q=-3.721 MeV) near the $N=$82
closed neutron shell yielded a reaction rate \citep{SFK98} nearly one order of magnitude
lower than the HF predictions \citep{MRO97}. The
calculations are based on a wide range of $\alpha$ potential
parameters derived from extensive $^{144}$Sm $\alpha$ elastic
scattering data. Measurements of $\alpha$ elastic scattering on
$^{92}$Mo \citep{FGM01} have been used to determine the
$^{96}$Ru($\gamma,\alpha$)$^{92}$Mo reaction rate
similar to the approach of \cite{MRO97}. This prediction depends
sensitively on the potential parameters and is considerably lower
than the HF calculations \citep{RaT00}. Similar
results have been obtained in $\alpha$ capture measurements on the
{\it p} nuclei  $^{96}$Ru \citep{RHH02} and $^{112}$Sn \citep{OMB02},
where the experimental rates are also much
lower than the HF predictions \citep{RaT00} for the
temperature range between 2.5 GK and 3.5 GK. Reasonably good agreement
with theory was only found in the measurement of the
($\alpha,\gamma$) cross section of $^{70}$Ge \citep{FKS96}. In
addition, recent (n,$\alpha$) experiments on $^{147}$Sm,
$^{143}$Nd, and $^{95}$Mo \citep{GKA00,KGA00,RKK03} 
have shown that present statistical models
overestimate the associated $\alpha$-induced
astrophysical rates by more than a factor of two. In the case of
$^{147}$Sm the $\alpha$-widths distribution of the
resonances showed indications of non-statistical effects
\citep{KGR04}.

In the following we want to study the impact of the nuclear
reaction rates on the {\it p}-process yields within the framework
of multi-mass zone simulations. After a short summary of the model
parameters we will discuss the results of the {\it p}-process
nucleosynthesis simulations. These were performed by modifying the
reaction rates within the uncertainty limits of the present
experimental studies and of the theoretical predictions. The
impact of these modifications will be analyzed in terms of the
time integrated {\it p}-process reaction flux and in terms of
abundance predictions for the {\it p} nuclei. A further goal of
the present work is to identify the most critical reaction rates
in the mass range $A>$57 with respect to the {\it p} abundances
and to identify the reactions, which need more detailed
experimental investigation.

\section{Model parameters and simulations}

The present nuclear reaction network comprises more than 20000
reactions connecting about 1800 nuclei from hydrogen to bismuth.
We have simulated the abundance evolution of the associated
isotopes in the framework of a parameterized type II SN shock
front model  \citep{RAH95}. The {\it p} process was investigated
in 14 different mass layers of the Ne/O burning zone of a 25
M$_{\odot}$ star for the temperature and density profiles proposed
by \citet{YoH02} (see table \ref{tab1}). 
These
profiles are comparable to the ones used by \citep{RAH95}.
It should be noted that a wide
range of progenitor masses are expected to contribute to the
p-process. We chose here a 25 M$_{\odot}$ star as a representative example,
in part to facilitate comparison with previous work, for example
\citet{CRZ00,WHW02}. \citet{RAH95} surveyed the type II supernova
$p$-process for a wide range of models and find only a weak 
dependence of the final $p$-process on progenitor mass. 

Fig.
\ref{rapp_fig2} shows the temperature and density profile of the
shock front passing through three of the layers reaching peak
temperatures of 2.4 GK, 2.6 GK, and 2.96 GK, respectively. Table
\ref{tab1} lists the mass fraction of the star enclosed by each
layer M$_{n}$ as well as the peak temperature and density reached
in each of the layers.

The final {\it p}-process abundances depend sensitively on the
choice of the initial seed abundance \citep{ArG03}. 
In the present
approach we, therefore, kept the seed abundances fixed in order
to study the sensitivity of the $p$ abundances to
nuclear reaction rates.
The seed
abundance distribution is determined by the $s$- and $r$-process
history of the stellar material at the time of star formation,
which is subsequently modified by in-situ nucleosynthesis during
the evolution of the star considered. The latter aspect is
particularly important since the Ne/O layers receive an abundance
contribution in the mass region $A\approx$70-90 from the weak $s$
process, which occurs during the preceding helium and carbon
burning phases \citep{KWG94,TEM00}. This $s$-process component
depends on the mass of the star but also on the neutron
irradiation provided by the $^{22}$Ne($\alpha$,n) neutron source.
There has been some speculation that the $^{22}$Ne($\alpha$,n)
source might be more efficient than previously anticipated
\citep{CRZ00}, which would cause a substantial enhancement in the
abundances of the {\it p}-only nuclei $^{92}$Mo and $^{96}$Ru . A
significantly enhanced $^{22}$Ne($\alpha$,n) reaction rate would,
however, result in an enormous overproduction of $s$-process
nuclei, in sharp conflict with the observed galactic abundance
patterns \citep{HWR02}. Also, new measurements of the
$^{22}$Ne($\alpha$,n) cross section at low energies \citep{JKM01}
and a detailed analysis of the $^{25}$Mg(n,$\gamma$)$^{26}$Mg
reaction channel \citep{koehler} provide stringent limits for the
uncertainty of this reaction rate \citep{KLU04} and exclude the
suggested enhancement factor. The seed abundances for the present
study (Fig. \ref{rapp_fig3}) were adopted from a pre-supernova
evolution model of a 25M$_{\odot}$ star \citep{RAH95,Ray02,ArG03}
to facilitate comparison with previous work.

For modeling the {\it p}-process nucleosynthesis yields all n-,
p-, and $\alpha$-induced capture rates as well as their inverse
photodisintegration rates were based on the HF predictions of the
NON-SMOKER code \citep{RaT00}. In addition, reaction rates on
light nuclei with $Z\le$8 were included from the work of
\citet{CF88}, \citet{WGG89}, \citet{RAC94}, and \citet{HHJ99} to
account for the impact of light particle capture reactions on the
proton, neutron, and alpha budget. The abundances of the following
35 {\it p} nuclei were investigated in the calculations:
$^{74}$Se, $^{78}$Kr, $^{84}$Sr, $^{92}$Mo, $^{94}$Mo, $^{96}$Ru,
$^{98}$Ru, $^{102}$Pd, $^{106}$Cd, $^{108}$Cd, $^{112}$Sn,
$^{113}$In, $^{114}$Sn, $^{115}$Sn, $^{120}$Te, $^{124}$Xe,
$^{126}$Xe, $^{130}$Ba, $^{132}$Ba, $^{136}$Ce, $^{138}$La,
$^{138}$Ce, $^{144}$Sm, $^{152}$Gd, $^{156}$Dy, $^{158}$Dy,
$^{162}$Er, $^{164}$Er, $^{168}$Yb, $^{174}$Hf, $^{180}$Ta,
$^{180}$W, $^{184}$Os, $^{190}$Pt, and $^{196}$Hg. 
Following 
the method described in \citet{RAH95} the mass
fraction X$_{i,n}$ of each {\it p} nucleus $i$ in each {\it
p}-process layer $n$ was calculated independently for all 14 layers.
The total mass $m_{i}$ of nucleus $i$ was determined by the
respective contributions weighted by the mass of each zone 
delimited by its two neighboring $p$-process layers,
\begin{equation}
m_{i}=\sum_{n \geq 1}^{n = 13} \frac{1}{2}(X_{i,n}+X_{i,(n-1)}) \times (M_{n}-M_{n-1}).
\end{equation}
with M$_{n}$ representing the mass of the star within 
layer $n$.

\subsection{{\it p}-process abundances}

The efficiency of a particular nucleosynthesis process in
contributing to the observed solar abundance distribution can be
expressed by the overproduction factor $\langle F_i\rangle$, which
compares the produced abundance of a given isotope $i$ with the
observed solar abundance. For the analysis of the present
calculations an overproduction factor has
been defined as
\begin{equation}
\langle F_{i}\rangle=\frac{m_{i}}{M_{\rm tot} X_{i0}}
\end{equation}
with the total mass 
$M_{\rm tot}=\sum_{n \geq
1}{M_{n}-M_{n-1}}=M_{13}-M_0$
 obtained by the sum over all {\it p}-process layers . The
solar abundance 
mass fractions $X_{i0}$ 
were taken from \citet{AnG89}. The averaged
overproduction factor for the 35 {\it p}-only nuclei is
\begin{equation}
F_{0}=\sum_{i}{\frac{\langle F_{i}\rangle}{35}}.
\end{equation}
The normalized overproduction factor $\langle F_{i}\rangle
/F_{0}$ is by definition equal to unity when the simulated
abundances match with the observed solar values \citep{RAH95}.

Fig. \ref{rapp_fig4} shows the normalized overproduction $\langle F_{i}\rangle$/F$_0$
as a function of mass number $A$. Given the uncertainties of the
present model and the uncertainties of the nuclear input the
calculated {\it p}-abundance pattern compares reasonably well
with the observed solar abundances. 
The comparison of the results displayed in Fig.~\ref{rapp_fig4}
with previous work relies on an
identical seed abundance distribution and on similar statistical
model based reaction rates. Within the expected model specific
discrepancies our calculations yield a p-nuclei abundance pattern,
which is very similar to the previous results.
Our prediction for $^{190}$Pt
disagrees with the result of \citet{RAH95} but is consistent with
the value given by \citet{ArG03}. This may reflect the use of
different reaction rate compilations.

The observed abundances of the {\it p} nuclei could be reproduced
in most cases within a factor of three.
Particular exceptions are the notorious
underproduction of $^{92,94}$Mo and $^{96,98}$Ru  and the
deficiencies of $^{113}$In, $^{115}$Sn, and $^{138}$La. 
 Also the much debated $^{180}$Ta abundance is found in agreement
 with previous calculations. However, this case has been treated
 without distinguishing between ground and isomeric state.
 Accordingly, the present value represents an upper limit for the
 $p$ contribution to that isotope, still compatible with the
 significant $s$-process contribution from AGB stars
\citep{KAH04}.

The low production of the light {\it p} nuclei $^{92,94}$Mo and
$^{96,98}$Ru has been observed in all previous {\it p}-process studies
\citep{RAH95,ArG03} and remains as the main enigma in {\it p}-process
simulations. Possible additional nucleosynthesis scenarios for
producing these isotopes have been discussed and
include a $\nu$-induced abundance component originating from the
neutrino-driven supernova shock \citep{HWF96}, the production in
the neutron-rich, alpha-rich freeze out near the mass cut in type
II supernovae \citep{Mey03}, and finally the independent production
mechanism by the {\it rp} process in the outer layers of accreting
neutron stars \citep{Sch98,DRM03}. Within the framework
of the {\it p}-process models possible explanations have
been related to modifications in the seed abundance
distribution \citep{CRZ00,ArG03}, as discussed before, or were suspected
to result from the nuclear physics parameters used
for calculating the associated reaction rates.

Also the underproduction of $^{113}$In, $^{115}$Sn, $^{138}$La,
$^{152}$Gd, and $^{164}$Er has been observed in previous calculations
\citep{RAH95,ArG03}. No other nucleosynthesis source has been
identified for $^{113}$In since it is effectively shielded from
the $s$- and $r$-process path \citep{NKT94,TKW98}. For the underproduced
$^{115}$Sn the $s$ process provides an additional 5\% production
through the sequence $^{114}$Cd(n,$\gamma$)$^{115}$Cd$^m$($\beta^-$)$^{115}$In$^m$
feeding the 1/2$^-$ isomeric state in $^{115}$In$^m$, which
subsequently decays through a weak $\beta^-$ branch to $^{115}$Sn
\citep{BWK89,NKT94}. More significant contributions to $^{113}$In and
$^{115}$Sn (12\% and 43\%, respectively) are expected from the $r$
process via the $\beta$-unstable isomers in $^{113}$Cd and $^{115}$In,
which are populated in the post-$r$-process decay chains \citep{NKT94,TKW98}.
Nevertheless, the origin of the rare In and Sn isotopes represents
an unsolved puzzle.

The abundances of $^{152}$Gd and of $^{164}$Er, however, are understood to
result mostly from $s$-process nucleosynthesis. In total 71\% of $^{152}$Gd are
produced by the $s$-process branching at $^{151}$Sm \citep{AAA04c}. The here
observed abundances for $^{152}$Gd originate mainly at relatively
cool temperatures T$_{9}<2.0$ typical for the outer
{\it p}-process layers. Similarly, $^{164}$Er is underproduced by the
{\it p} process but more than 90\% can be explained by feeding through
the high temperature $s$-process branching via bound-state $\beta$-decay of
$^{163}$Dy to $^{163}$Ho with the subsequent neutron capture sequence
$^{163}$Ho(n,$\gamma$)$^{164}$Ho($\beta^-$)$^{164}$Er
\citep{JaK96,BSA01}.

The nucleus $^{138}$La is systematically underproduced in all
supernova based {\it p}-process models \citep{ArG03}. Alternative
models related to energetic stellar particles \citep{HSB76} or
neutrino interactions \citep{WHH90} have been developed to explain
the observed $^{138}$La abundances. A detailed analysis of the
$^{138}$La problem has been reported by \cite{GAB01}. In terms of
nuclear reaction rates it was shown that plausible estimates for
the related uncertainties can not explain the underproduction of
$^{138}$La. This uncertainty could be removed, however, if the
theoretical production rate $^{139}$La($\gamma$,n) and the
destruction rate $^{138}$La($\gamma$,n) could be replaced by
experimental data.

The final {\it p} abundances depend on
the {\it p}-process reaction flux and reaction branchings, which
determine the feeding and depletion of the various {\it p} nuclides.
In the following section we discuss the characteristic
flux patterns as a function of mass layer and/or peak
temperature.

\subsection{{\it p}-process flux}

The time integrated reaction flux per mass layer provides
information about the main reaction path during the
nucleosynthesis event and serves as a tool for monitoring the
effects of nuclear structure parameters such as shell closure or
deformation on reaction path and reaction branchings. The time
integrated reaction flux corresponds to the net number of
reactions between two isotopes $i$ and $j$ integrated over a
certain period of time $\Delta t = t_1 - t_0$. This flux is defined
by
\begin{equation}
f_{i,j}=\int_{t_0}^{t_1}\left[\frac{dY_i}{dt}_{(i\rightarrow j)}
-\frac{dY_j}{dt}_{(j\rightarrow i)}\right]dt \label{E2}
\end{equation}
\noindent with the isotopic abundances $Y_i=X_i/A_i$ (mass
fraction divided by mass number). 
In the following we refer to the time integrated reaction flux
simply as reaction flux.
The maximum reaction flux $f_{i,j}$
defines the main reaction path along which nucleosynthesis will
take place. 

\begin{figure}[htp]
\plottwo{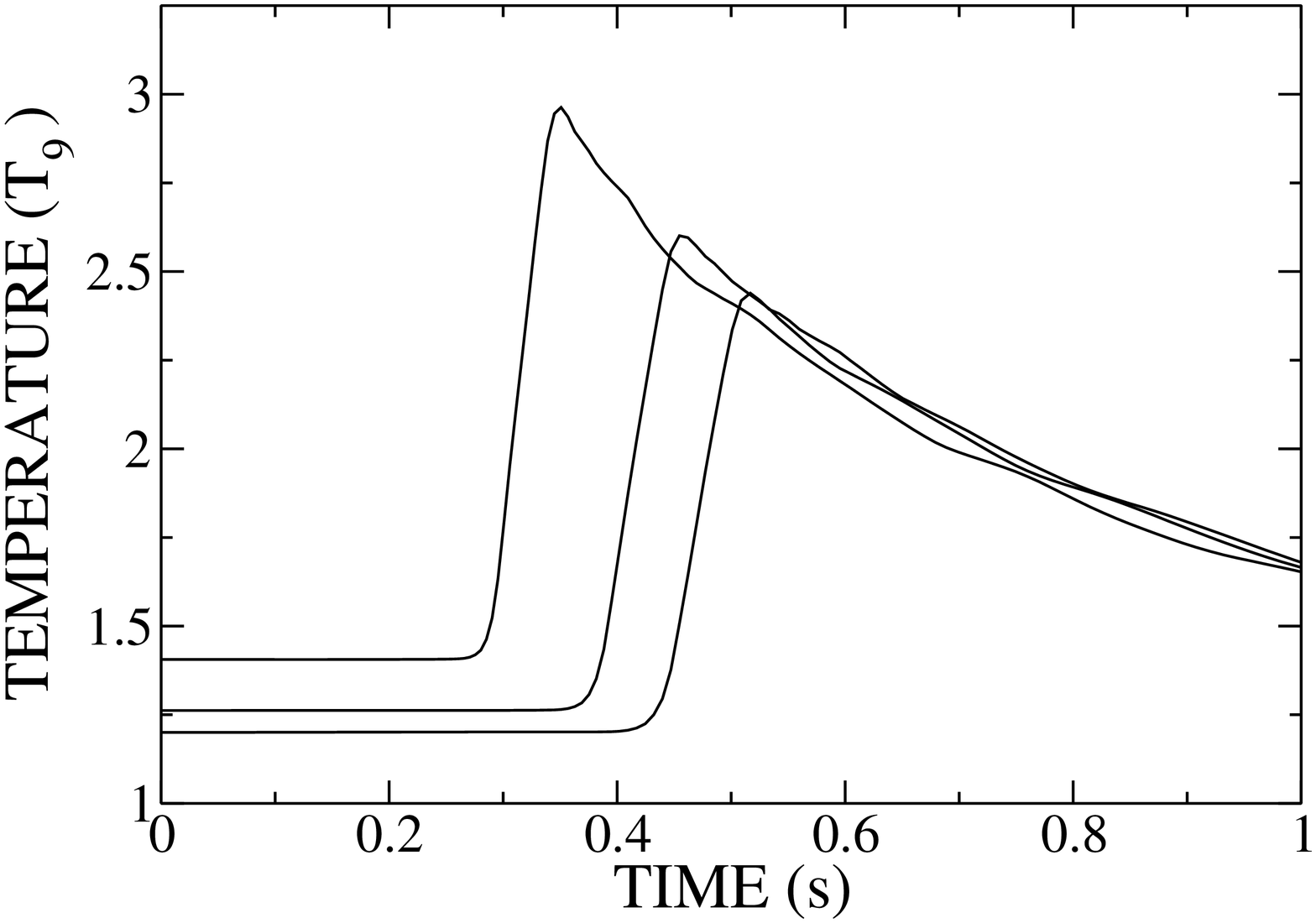}{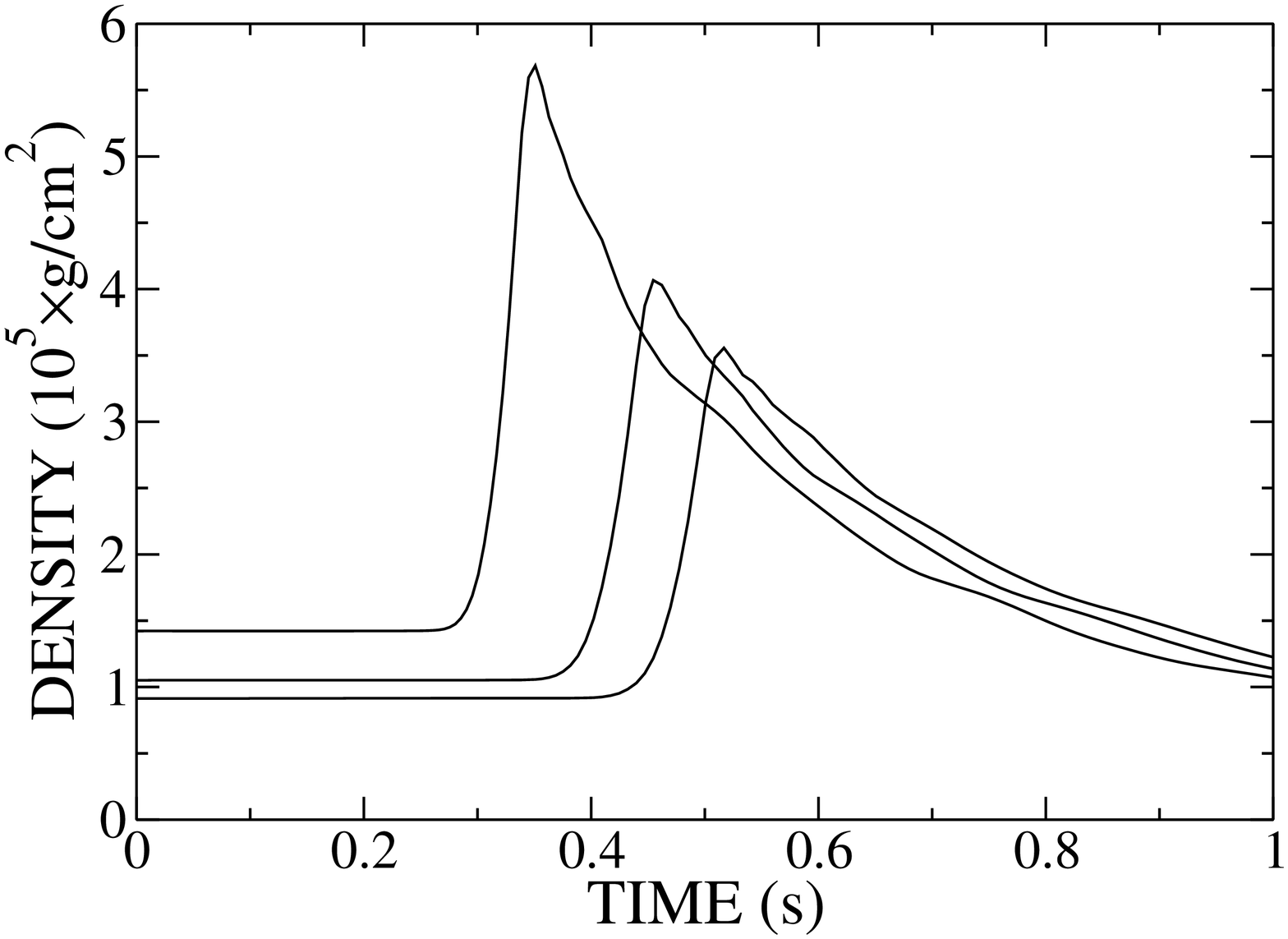}
\caption{Temperature and density profiles of the supernova shock front traversing
the Ne/O layer of the pre-supernova star.\label{rapp_fig2}}
\end{figure}

\begin{figure}
\epsscale{.80} \plotone{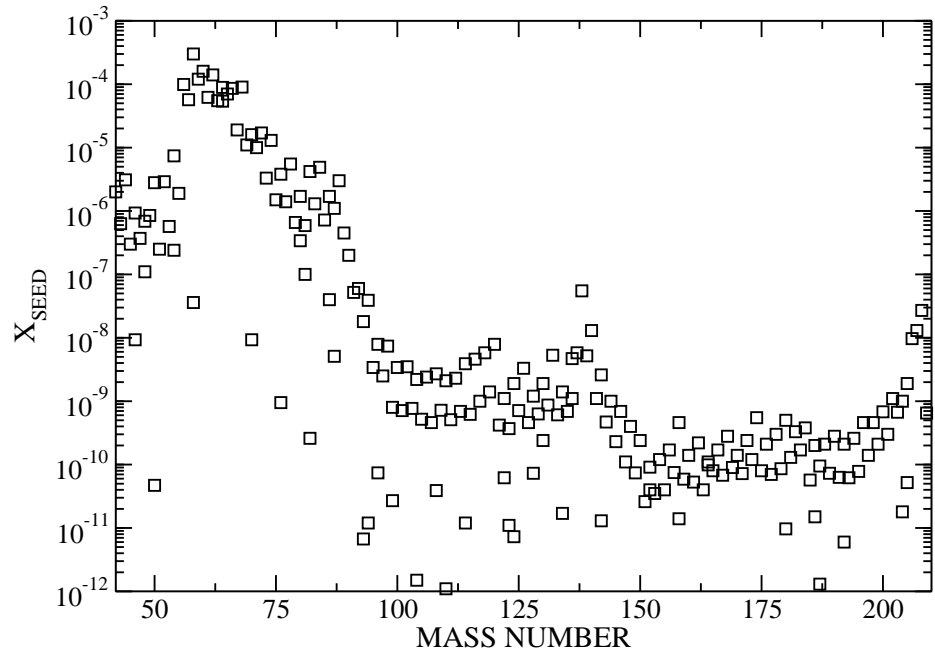}
\caption{Adopted seed abundance distribution for the {\it p}-process simulations.
\label{rapp_fig3}}
\end{figure}

\begin{figure}
\epsscale{.70} \plotone{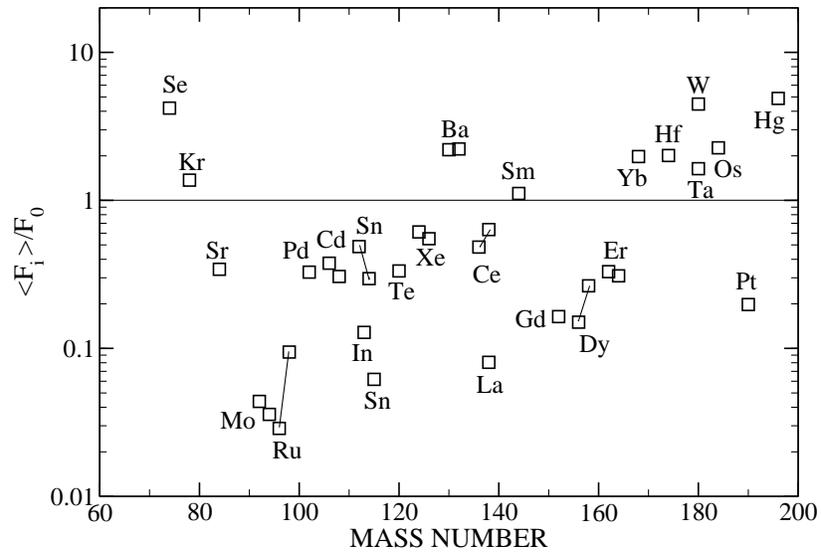}
\caption{Averaged normalized overproduction factor for the proton rich {\it p} nuclei from
network calculation with standard reaction rates (see text).
\label{rapp_fig4}}
\end{figure}

\begin{figure}
\epsscale{1.0}
\plotone{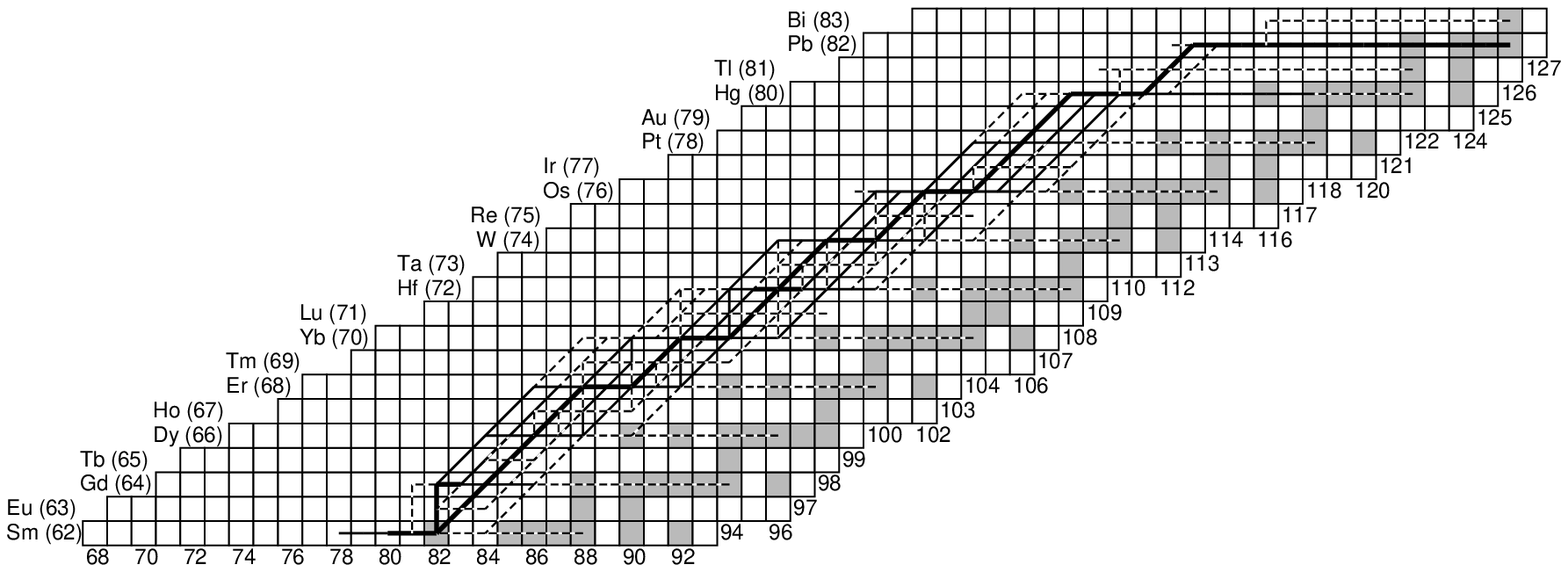}
\plotone{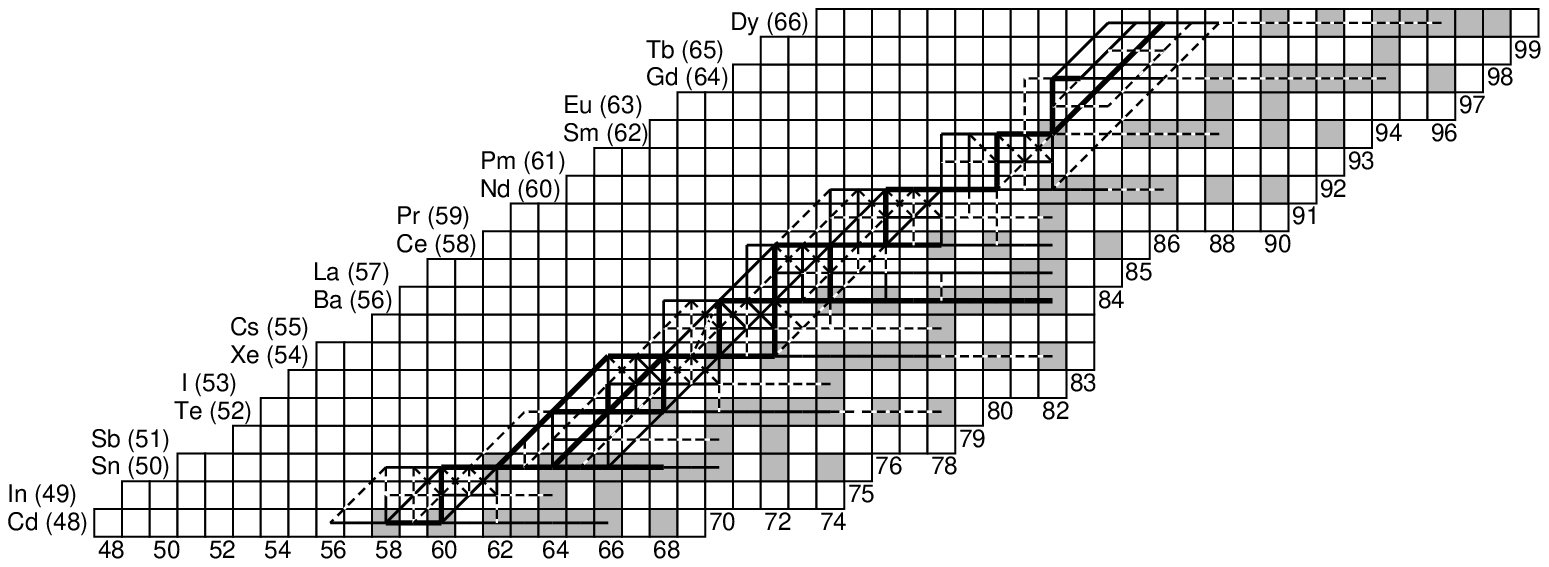}
\plotone{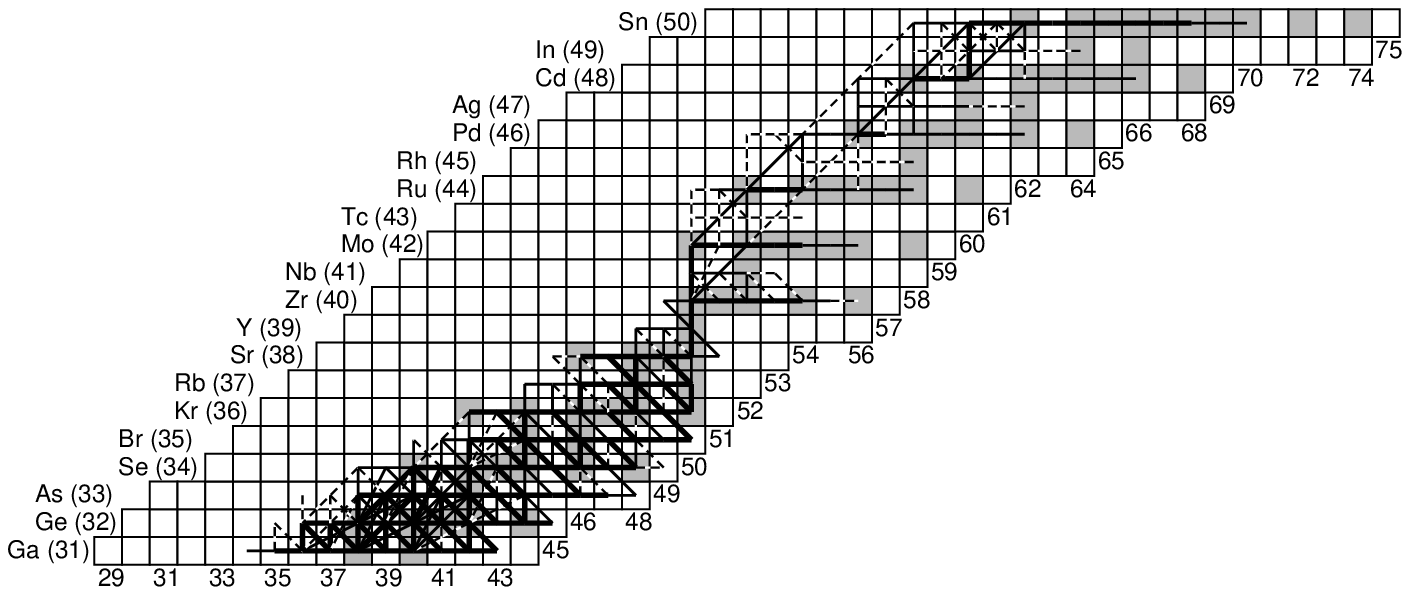} \caption{Integrated reaction
flux during the first second of a type II supernova explosion in
the Ne/O layer with a maximal temperature of T$_{9}$=2.96. The
reaction flux is
indicated by the line thickness (thick solid line $>$10$^{-10}$,
solid line $>$10$^{-11}$, dashed line
$>$10$^{-12}$).\label{rapp_fig5}}
\end{figure}

\begin{figure}
\epsscale{1.0} \plotone{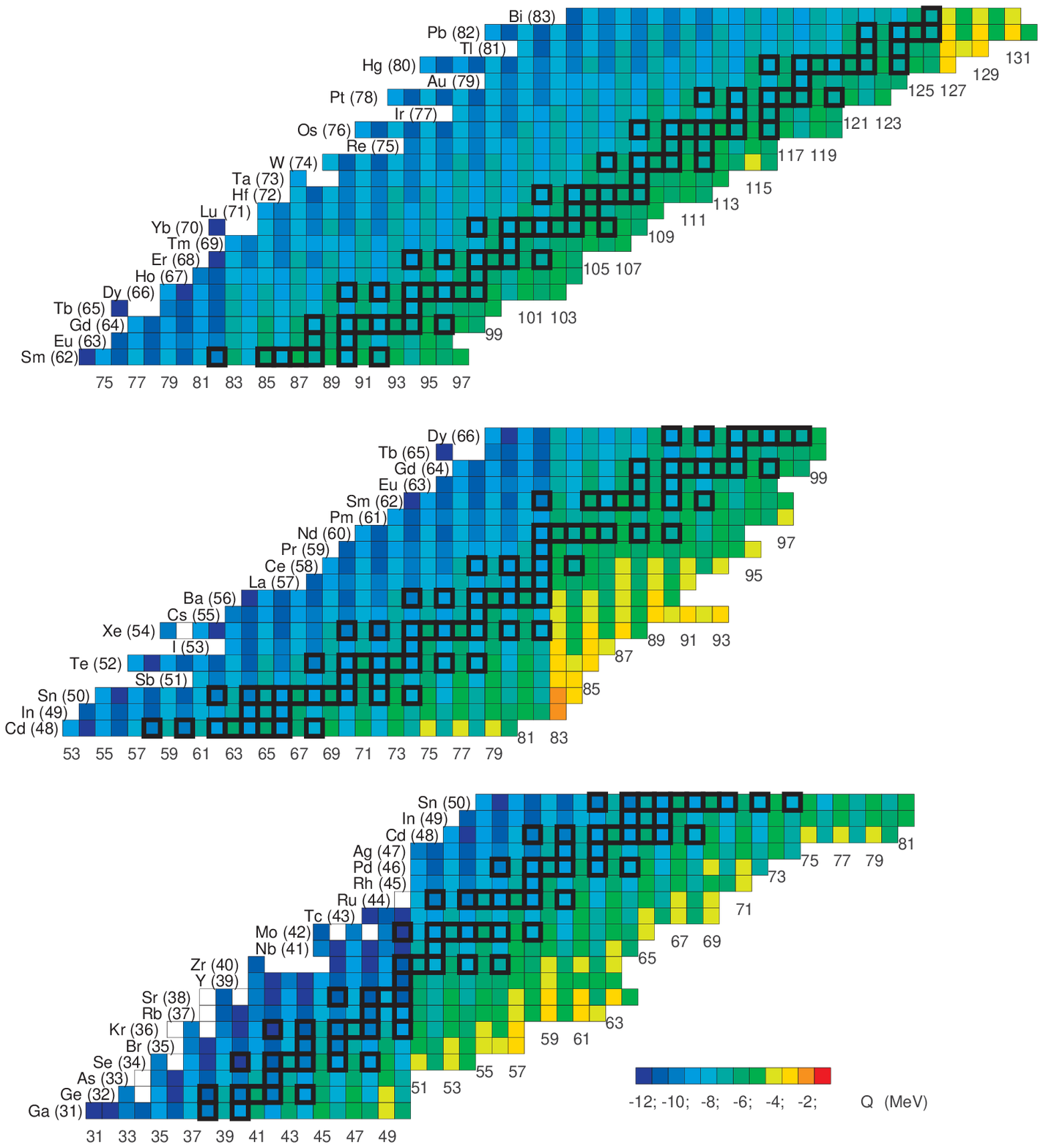}
\caption{Q-values for ($\gamma$,n) reactions based on experimental mass data
\citep{AWT03}.\label{rapp_fig6}}
\end{figure}

\begin{figure}
\epsscale{1.0} \plotone{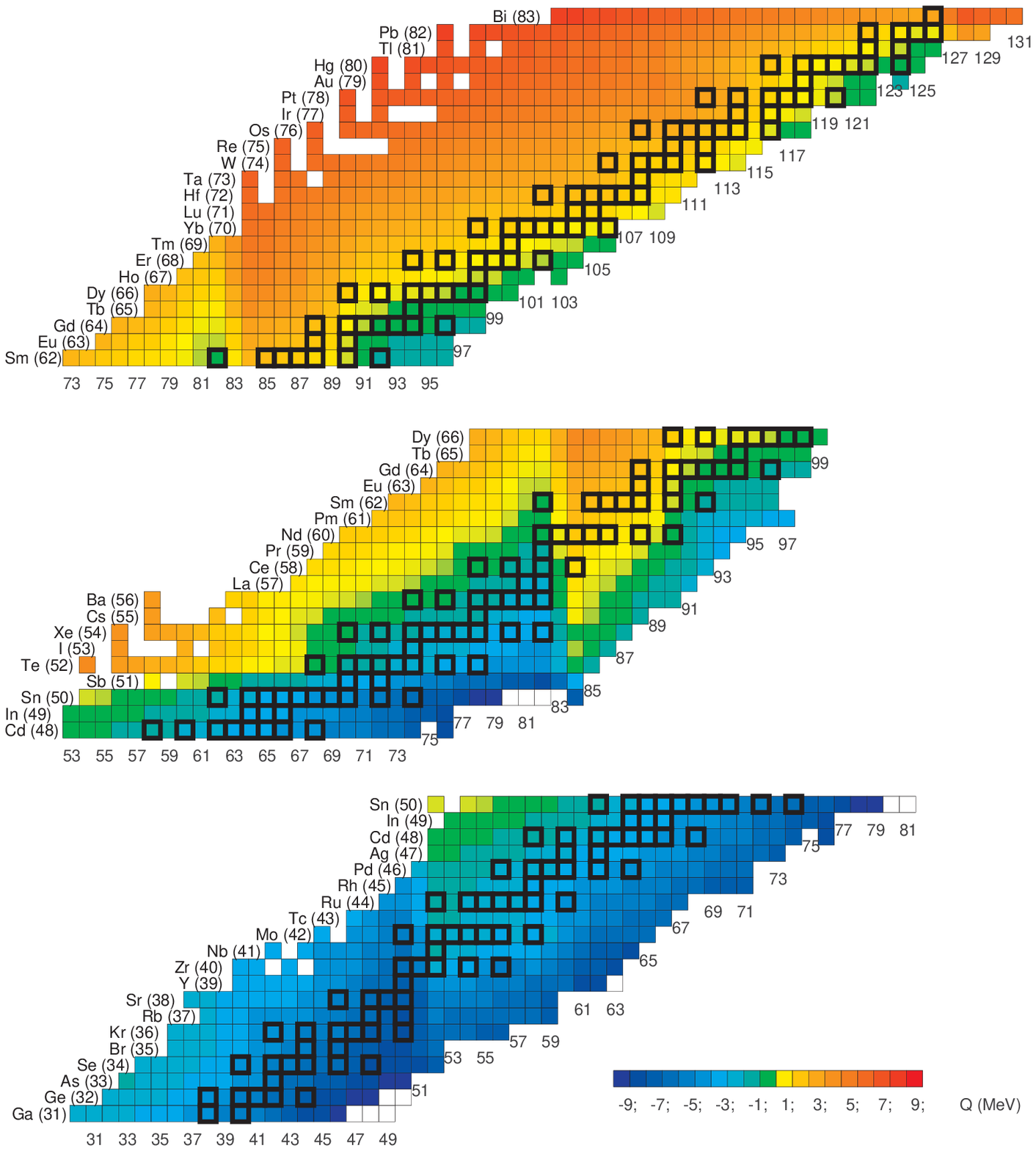}
\caption{Q-values for ($\gamma$,$\alpha$) reactions based on experimental mass
data \citep{AWT03}. \label{rapp_fig7}}
\end{figure}

The reaction flux has been calculated separately for each investigated
layer and depends strongly on the respective peak temperatures. 
The temperature and density profiles are shown in Fig. \ref{rapp_fig2}. 
Fig. \ref{rapp_fig5} shows the reaction flux integrated 
over a 1~s time interval covering the entire shock-driven temperature
peak of T$_9$=2.96 shown in Fig.~\ref{rapp_fig2}
During this period the seed nuclei in the
range 62$<Z<$83 are processed by ($\gamma$,n) reactions towards
the neutron deficient side of stability (upper part). With
increasing neutron binding energy of the reaction products (Fig.
\ref{rapp_fig6}) the abundance distribution is driven towards
lower masses by the competing ($\gamma$,$\alpha$) channel. Because
the neutron deficient isotopes above $N=$82 are $\alpha$-unbound as
shown in Fig. \ref{rapp_fig7}, spontaneous decay is only inhibited
by the high Coulomb barriers. For neutron magic nuclei the rapid
increase in neutron and $\alpha$ binding energy reduces the
($\gamma$,n) and ($\gamma,\alpha$) photodisintegration rates
significantly and forces the reaction flux into the ($\gamma$,p)
channel along the $N=$82 closed shell nuclei towards lower masses.
In particular the {\it p}-process isotope $^{144}$Sm is formed by
feeding through the $^{147}$Eu($\gamma$,p) and the
$^{148}$Gd($\gamma,\alpha$) channels, but is depleted through
$^{144}$Sm($\gamma$,n) photodisintegration.

The middle part of Fig.~\ref{rapp_fig5} shows the integrated
reaction flux for 48$<Z<$62. While in this mass range the
overall reaction path is still driven by ($\gamma$,n) reactions
towards the neutron deficient side of stability, the flux pattern
is characterized by a strong ($\gamma$,p) reaction component
towards stability for $N<$82 since the rapid increase in $\alpha$
binding energy (Fig. \ref{rapp_fig7}) inhibits the
($\gamma$,$\alpha$) channel. The ($\gamma$,$\alpha$) processes
regain a more competitive role for nuclei with $Z<$58 and provide
a strong reaction flux towards the even-even proton magic Sn
isotopes with $Z=$50. The main reaction channels pass through
$^{112}$Sn and $^{114}$Sn, which are both fed via ($\gamma$,n) and
($\gamma,\alpha$) reactions. 

The {\it p}-process flux below $Z=$50 (lower part of Fig.
\ref{rapp_fig5}) is still characterized by ($\gamma,\alpha$)
reactions feeding the {\it p} nuclei $^{106,108}$Cd, $^{102}$Pd
and eventually $^{96}$Ru and $^{92}$Mo. At $N=$50 the
($\gamma,\alpha$) flux is largely diminished because the
increasing $\alpha$ binding energy efficiently reduces the
($\gamma,\alpha$) channel. An additional strong ($\gamma$,n) and
($\gamma$,p) flux occurs towards $N=$50 closed neutron shell nuclei,
which is followed by ($\gamma$,p) photodisintegration along the
$N=$50 isotone chain. Compared to the integrated flux in the region
$Z\ge$50 the overall flux is significantly reduced because of the
high single particle binding energies in this region near the
$Z, N=$50 closed shells. The figure indicates that the abundances of
the underproduced {\it p} nuclei such as $^{92}$Mo and $^{96}$Ru
are mainly determined by the ($\gamma$,n) reaction flux feeding
and the ($\gamma$,p) flux depleting these isotopes. In addition 
there is a significant flux through the $^{96}$Ru($\gamma,\alpha$)$^{92}$Mo
reaction. The final abundances of the
{\it p} nuclei $^{94}$Mo and $^{98}$Ru depend mainly on the
($\gamma$,n) rates depleting these isotopes with $^{94}$Mo being
mainly produced at the lower temperature mass zones. Only a few
($\gamma,\alpha$) rates contribute to the reaction flux at $Z<$50,
mainly due to the rapid increase in $\alpha$-binding energy as
shown in Fig. \ref{rapp_fig7}.

In the mass range below the $N=$50 neutron shell the nuclei become
more and more resistant against photodisintegration because of
their high neutron and proton binding energies (Figs.
\ref{rapp_fig6} and \ref{rapp_fig8}). The reaction flux is
characterized by a complex pattern of single neutron, and proton
capture and their inverse dissociation reactions. These processes
are accompanied by (p,n) and inverse (n,p) reactions because of
their typically low
thresholds
of $\approx$ 0.5 - 1.0 MeV. The
reaction path remains close to stability.

A case of {\it p}-process nucleosynthesis in a layer reaching only
a peak temperature of T$_{9}$=2.44 is displayed in Fig.
\ref{rapp_fig9}. For isotopes with 62$<Z<$82, shown in the upper
part of the figure, the reaction flux is dominated by ($\gamma$,n)
reactions, which drive the initial seed abundance distribution
towards the neutron deficient side of the line of stability. After
the shock front passed, the produced radioactive isotopes decay
back to the line of stability. Only minor changes are anticipated
for the overall abundance distribution in this mass range. Besides
the ($\gamma$,n) reactions a few ($\gamma$,$\alpha$) reactions can
be observed processing neutron deficient isotopes in the Hf to Pt
range towards lower masses. At these temperatures the production
of $^{162}$Er and $^{184}$Os (Fig. \ref{rapp_fig10}) is limited to
the ($\gamma$,n) reactions along the $Z=$68 and $Z=$76 isotope chains,
respectively. In the case of $^{164}$Er, no feeding through
$^{168}$Yb($\gamma,\alpha$)$^{164}$Er seems to be taking place.
The {\it p} contribution to $^{152}$Gd depends on the ($\gamma$,n)
reactions along the $Z=$64 isotope chain, but is further depleted
through $^{152}$Gd($\gamma,\alpha$)$^{148}$Sm. The $N=$82 closed
shell isotope $^{144}$Sm is mainly fed by
$^{148}$Gd($\gamma,\alpha$)$^{144}$Sm reactions.

In this mass layer the time integrated flux between the $N=$82 and $Z=$50
closed shells is again dominated by ($\gamma$,n) reactions
subsequently balanced by inverse (n,$\gamma$) neutron capture
reactions as shown in Fig. \ref{rapp_fig9}. Therefore, the abundance distribution
changes only along the isotopic chain. Both ($\gamma$,p)
and ($\gamma,\alpha$) reactions are negligible. The {\it p} nuclei in
this mass range are not affected by the reaction flux with two
exceptions, $^{138}$La and $^{138}$Ce. This pattern continues also
towards lower masses down to the $N=$50 neutron closed shell. The
($\gamma$,n) reaction flux associated with the $^{92,94}$Mo and
$^{96,98}$Ru nuclei is negligible because of the high neutron
binding energies. These results suggest that the {\it p}-process production
of these isotopes is confined to higher temperature layers.

For nuclei below the $N=$50 closed neutron shell the reaction
pattern differs distinctively from the higher mass regions.
Similar to the flux at higher temperatures, photodisintegration
into the neutron and proton channel and their inverse capture
processes as well as (p,n) and inverse (n,p) reactions define the
reaction path and determine the final abundance distribution in
the mass range below $N=$50.

As shown in Figs. \ref{rapp_fig5} and \ref{rapp_fig9} the
integrated {\it p}-process reaction flux is considerably different
between high (T$_9$=2.96) and low temperature (T$_9$=2.4) layers.
In particular the nucleosynthesis of higher mass {\it p} nuclei
such as $^{156}$Dy, $^{162}$Er, $^{164}$Er, $^{168}$Yb,
$^{174}$Hf, $^{180}$W, $^{184}$Os, and $^{190}$Pt is temperature
sensitive because of the strong temperature dependence in the
($\gamma,\alpha$) reaction rates, which determine the reaction
flux pattern. The critical peak temperature for these reactions is
T$_9$=2.6, where the reaction path is still largely characterized
by a strong ($\gamma,\alpha$) flux towards lower mass isotopes.
Below T$_9$=2.6 the ($\gamma,\alpha$) reaction rates are typically
too weak to warrant rapid depletion of high $Z$ isotopes. This
causes an enrichment of the {\it p} nuclei through ($\gamma$,n)
feeding from the initial seed abundance distribution. For
temperatures above 2.6 GK the abundances of the heavy {\it p}
nuclei are reduced due to significant ($\gamma,\alpha$) depletion,
which increases exponentially with temperature.

The reaction flux depends on the initial seed abundance as well as
on the reaction rates. It has been argued \citep{ArG03} that the
reaction rates have only a limited influence on the final
{\it p}-process abundance distribution and in particular will not solve
the mystery of the light {\it p}-nuclei abundances. In the following we
seek to investigate in an independent systematic study
the direct impact of {\it p}-process reaction rates.

\section{The influence of reaction rates}

The influence of reaction rates on the final {\it p}-process
abundance distribution was studied first by introducing global
enhancement and reduction factors for different reaction channels
and by probing the sensitivity of the different ($\gamma$,n),
($\gamma,\alpha$), and ($\gamma$,p) photodisintegration branches.
All other input parameters and conditions remain unchanged in the
model. The approach of collectively changing all rates helps to
identify the mass range where rates have a direct impact on the
{\it p}-process flux and {\it p}-process abundance predictions.
The final p-abundances are sensitive to reaction rates
because they determine the balance of feeding and depletion flows
for each p-nucleus. In addition, changes in reaction rates change
branching ratios when several reaction channels compete, for
example ($\gamma$,n) with ($\gamma$,p) in the element range up to
$Z=64$ (see Fig. 5). In addition, such branchings are sensitive to
nuclear structure effects such as shell closures.

\subsection{Collective change of neutron- proton- and $\alpha$-rates}

As a first test of the sensitivity of the {\it p}-abundance distribution the
reaction rates for all neutron-induced processes and their inverse
photodisintegration reactions on nuclei with $A>$57 were enhanced
or reduced by a factor of three. In each case the {\it p}
abundances were calculated and the overproduction factors were
compared with the results based on the standard set of HF rates
used in this study \citep{RaT00}. Sensitivity
studies for proton- and $\alpha$-induced rates were performed in a similar
way. Fig. \ref{rapp_fig10} shows the respective overproduction factors
calculated with the modified rates in comparison with the ones
obtained with the standard HF rates.

The impact of the collective change of all (n,$\gamma$) and of the
inverse ($\gamma$,n) rates on the {\it p}-nuclei abundances is
shown in the top part of Fig. \ref{rapp_fig10}. One observes 
a correlation
over the entire mass range between the rates
for the neutron channel and the {\it p}-process yields. This is
not too surprising because ($\gamma$,n)-reactions are active in
all of the investigated layers where the {\it p} process is taking
place. The overall abundances of the {\it p} nuclei change by less
than a factor of two on average. Since the reaction flux below
$N=$50 is driven by capture and photodisintegration reactions the
abundances of {\it p} nuclei with $A\le$80 correlate more strongly
with changes of the reaction rate. A more pronounced correlation
can also be observed for $^{162}$Er, $^{164}$Er, and $^{168}$Yb,
which are sensitive to the ($\gamma$,n) photodisintegration rates
in the mass layers exposed to lower peak temperatures 2.76 $\leq
T_9 \leq 2.32$ as indicated for the example of $^{162}$Er in Fig.
\ref{rapp_fig11}.

\begin{figure}
\epsscale{1.0} \plotone{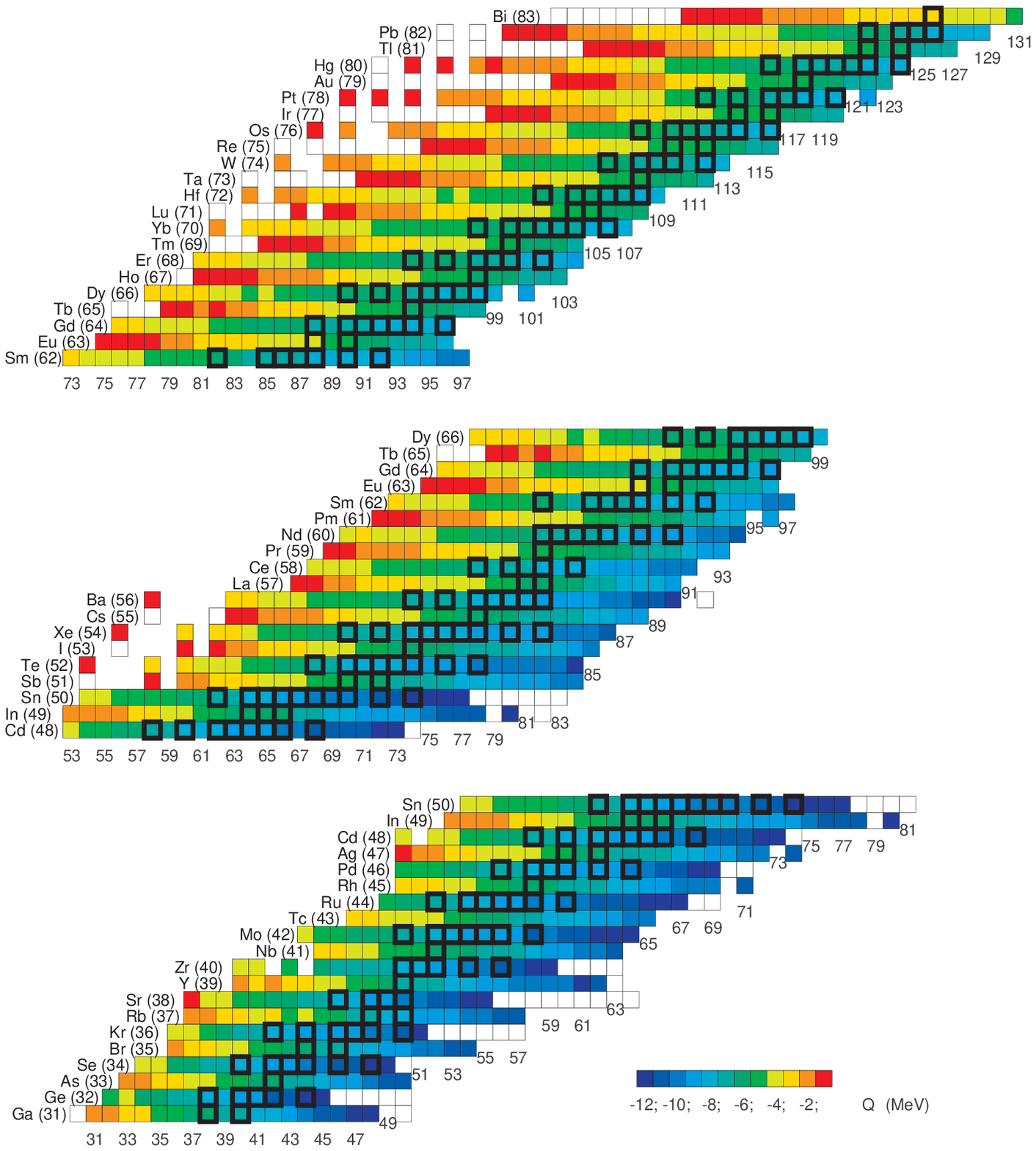}
\caption{Q-values for ($\gamma$,p) reactions based on experimental mass data
\citep{AWT03}.\label{rapp_fig8}}
\end{figure}

\begin{figure}
\epsscale{1.0} \plotone{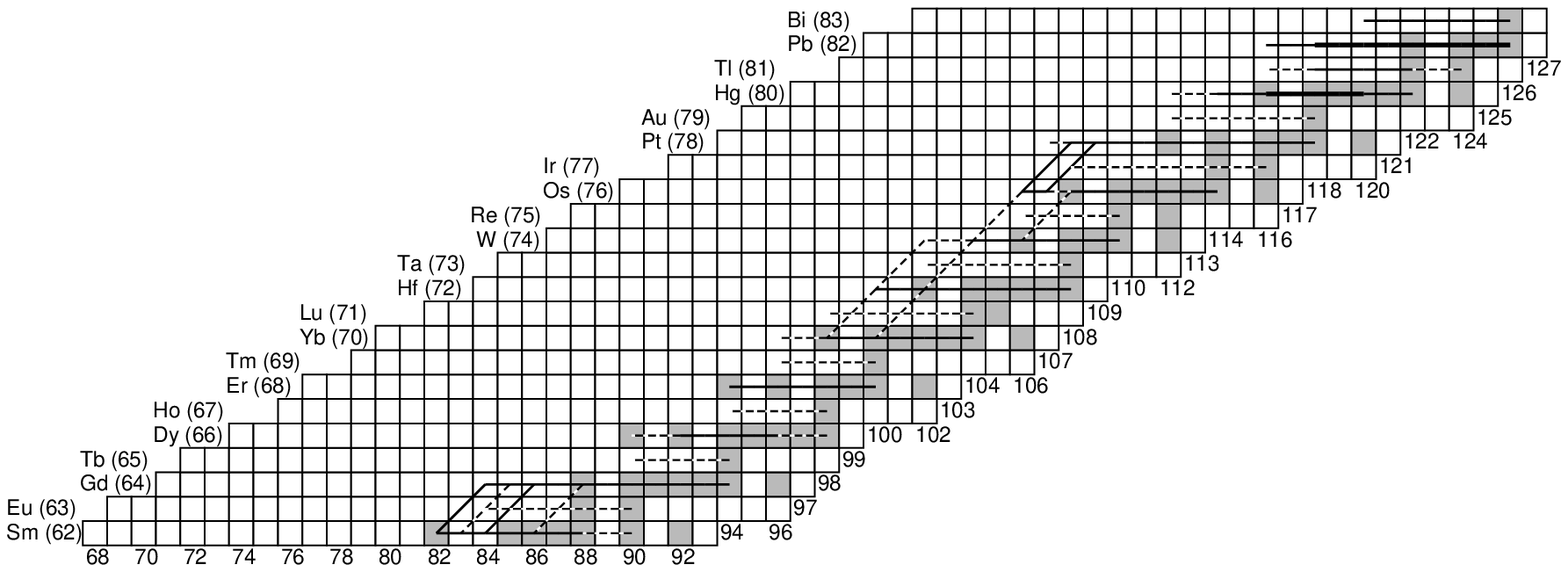}
\plotone{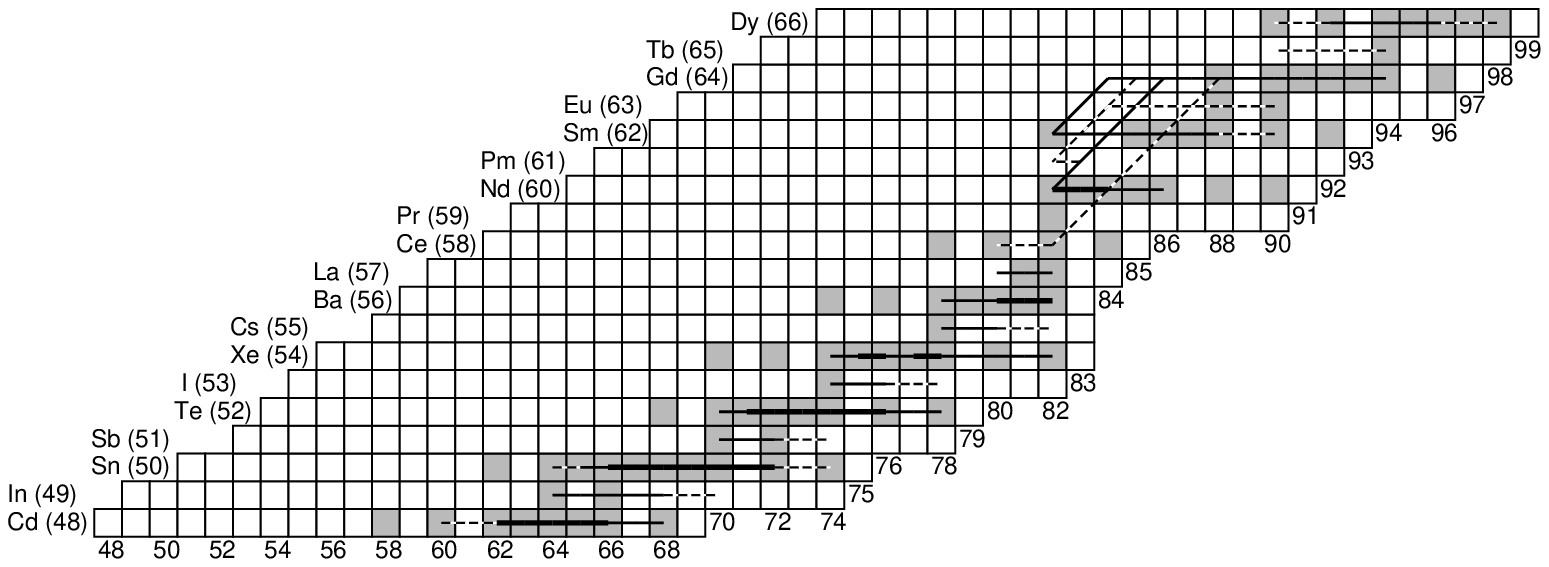}
\plotone{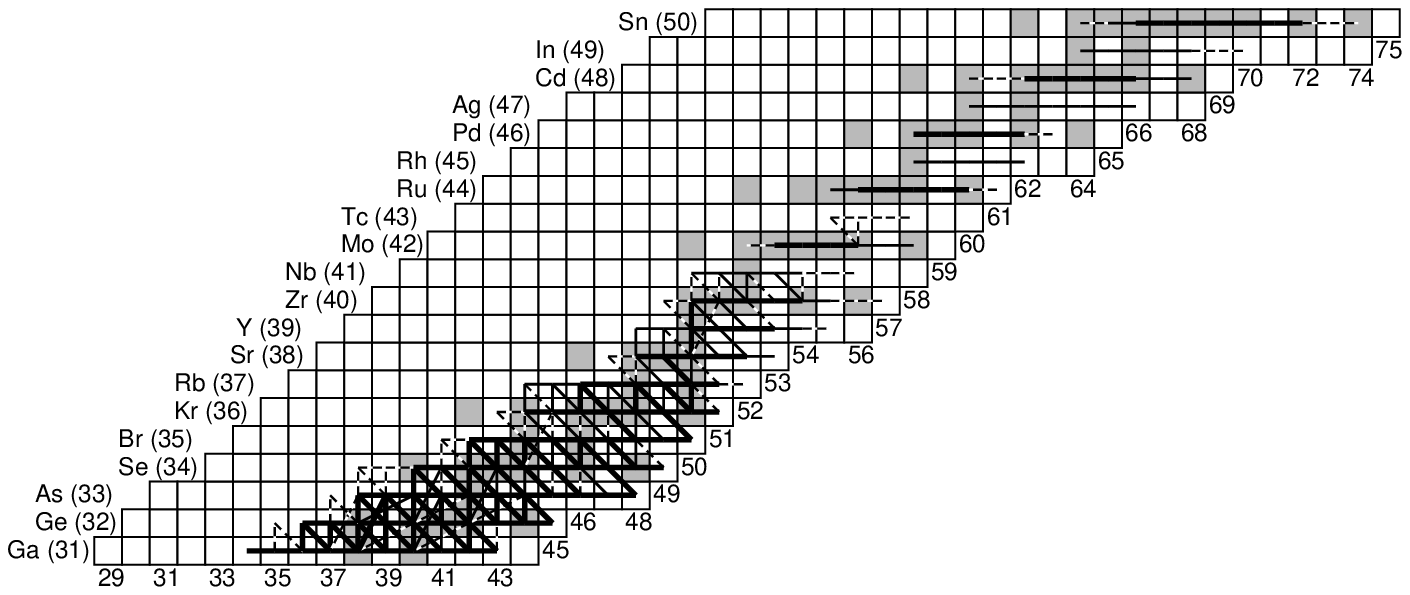} \caption{The integrated
reaction flux during the first second of a type II supernova
explosion in the Ne/O layer with a maximal temperature of
T$_{9}$=2.44. The reaction flux
is indicated by the line thickness
(thick solid line $>$10$^{-11}$, solid line $>$10$^{-12}$, dashed
line $>$10$^{-13}$).\label{rapp_fig9}}
\end{figure}

\begin{figure}[tp]
\begin{center}
\includegraphics*[width=8.5cm]{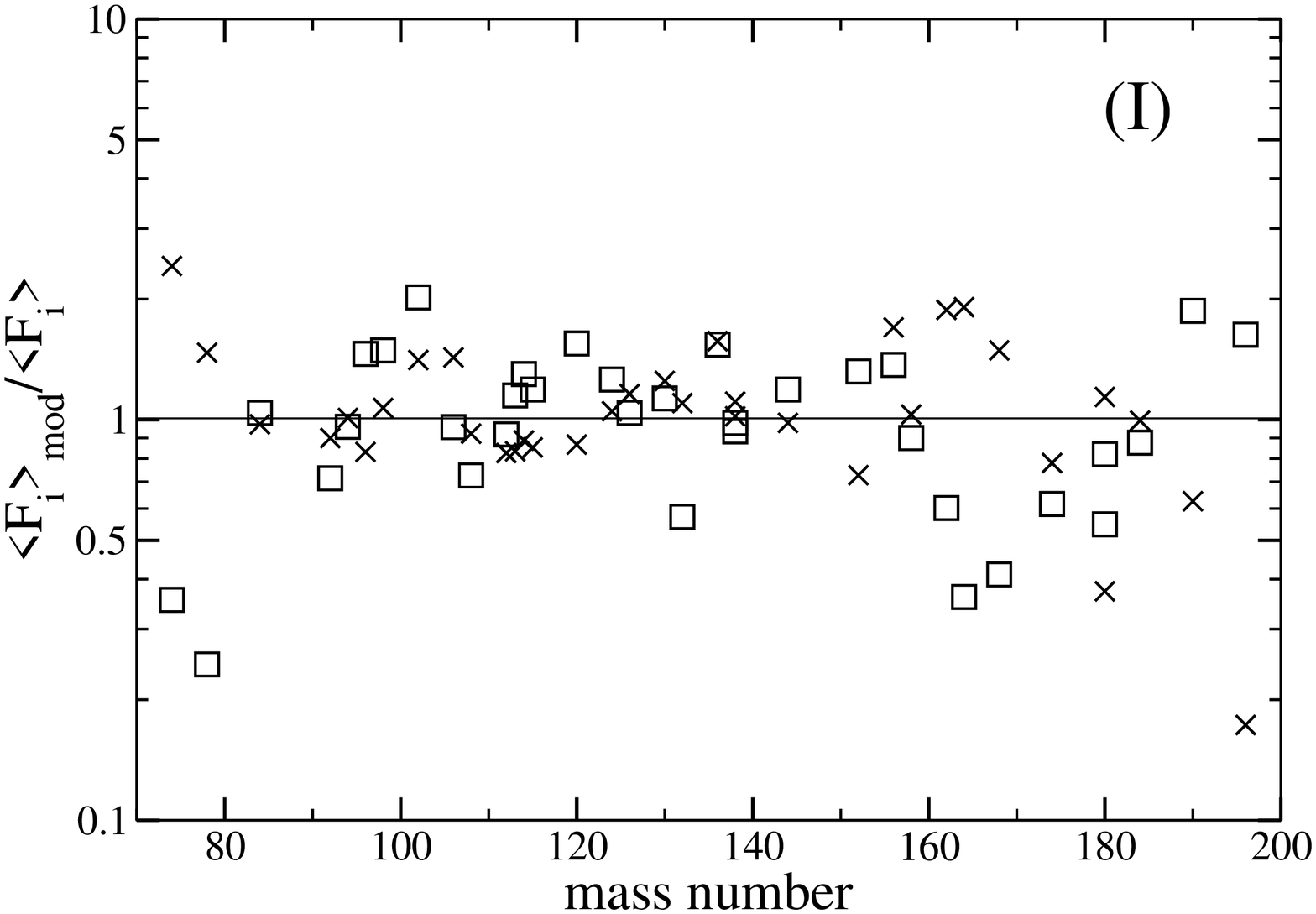}
\includegraphics*[width=8.5cm]{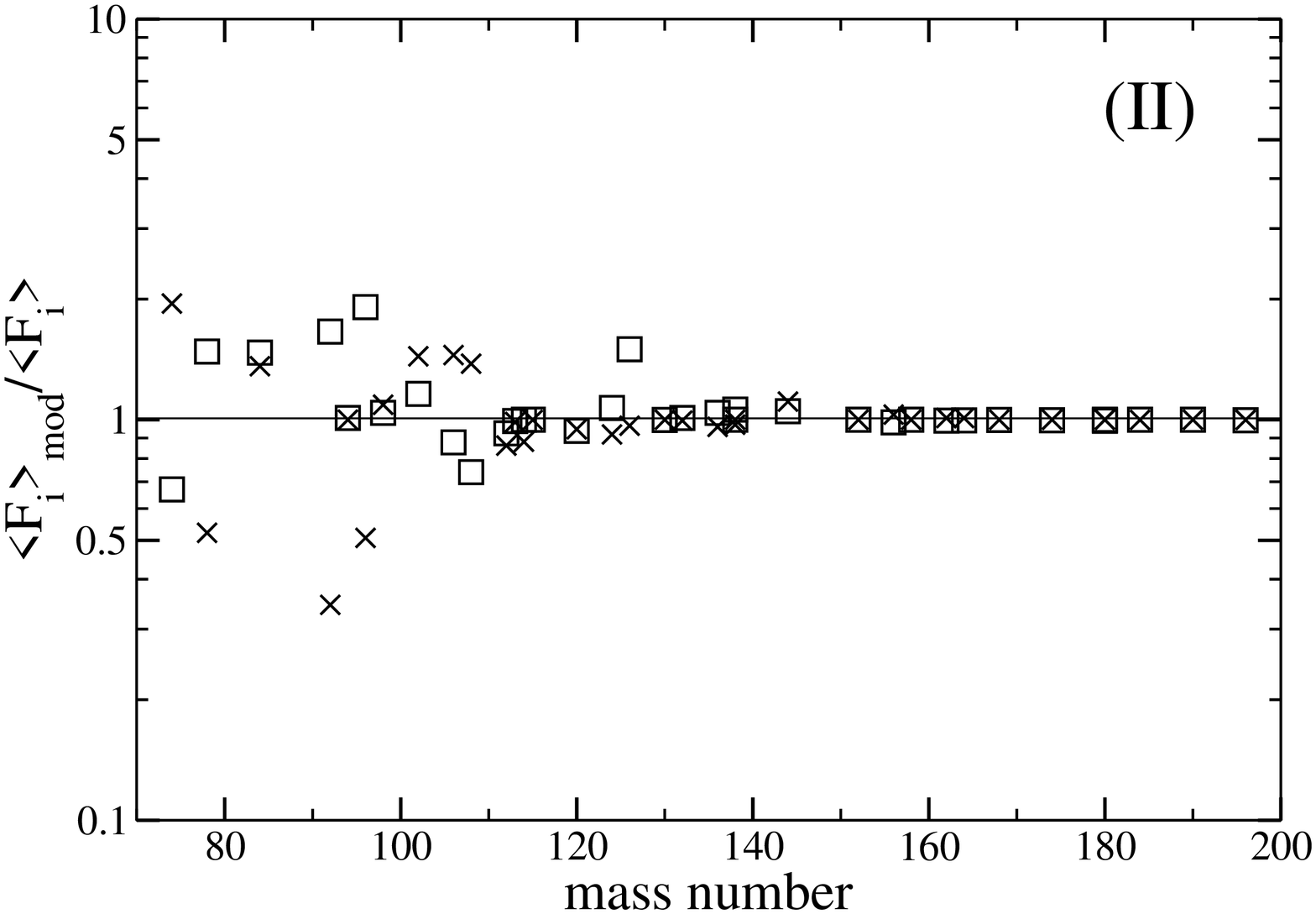}
\includegraphics*[width=8.5cm]{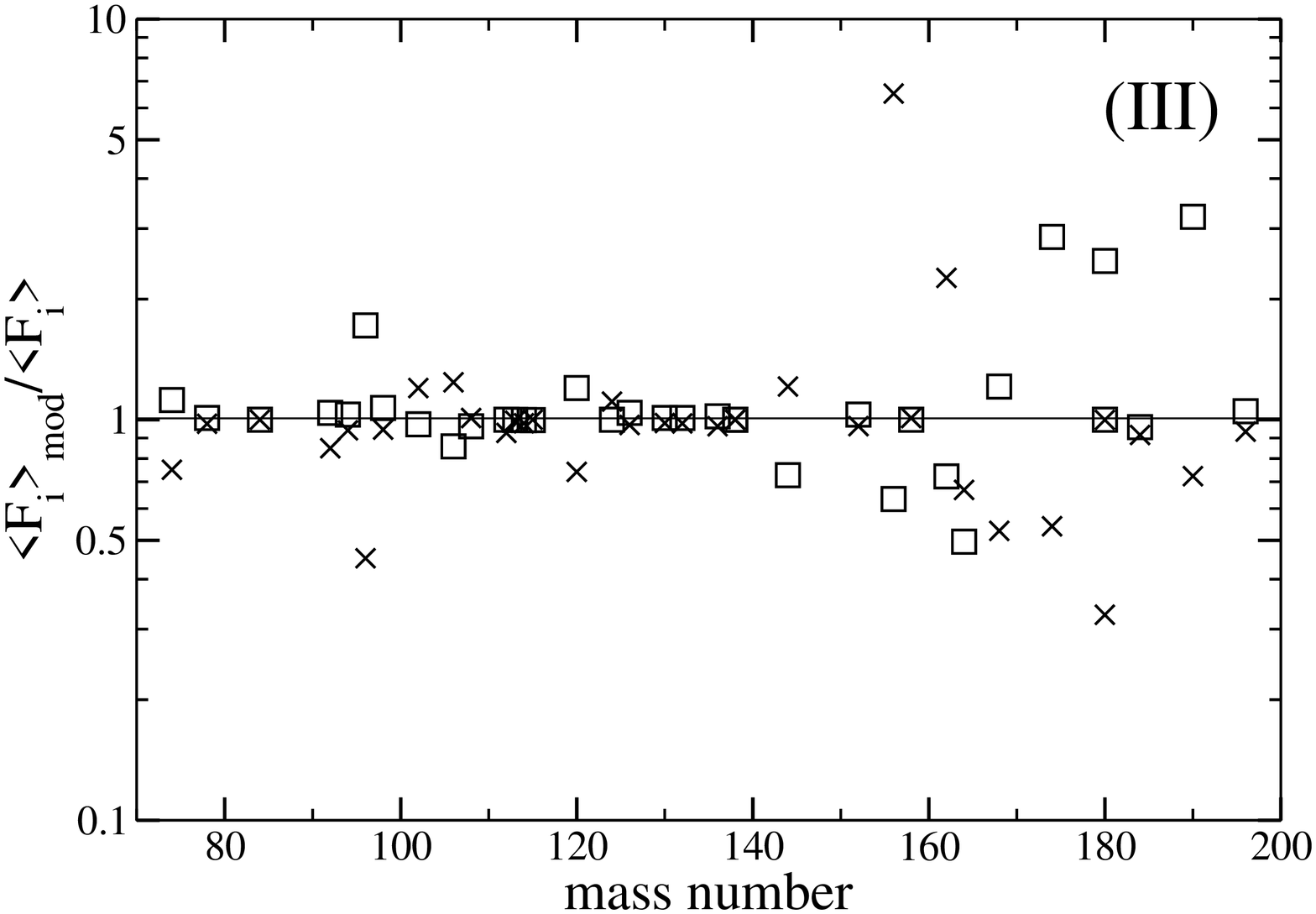}
\caption{Ratio of {\it p} abundances calculated with modified rates
and the presently accepted HF-rates for all n-induced (top), p-induced (middle),
and $\alpha$-induced reactions (bottom) and their inverse processes. 
Squares and crosses denote results obtained
with three times smaller and larger rates, respectively.
\label{rapp_fig10}}
\end{center}
\end{figure}

\begin{figure}
\begin{center}
\includegraphics*[width=9cm]{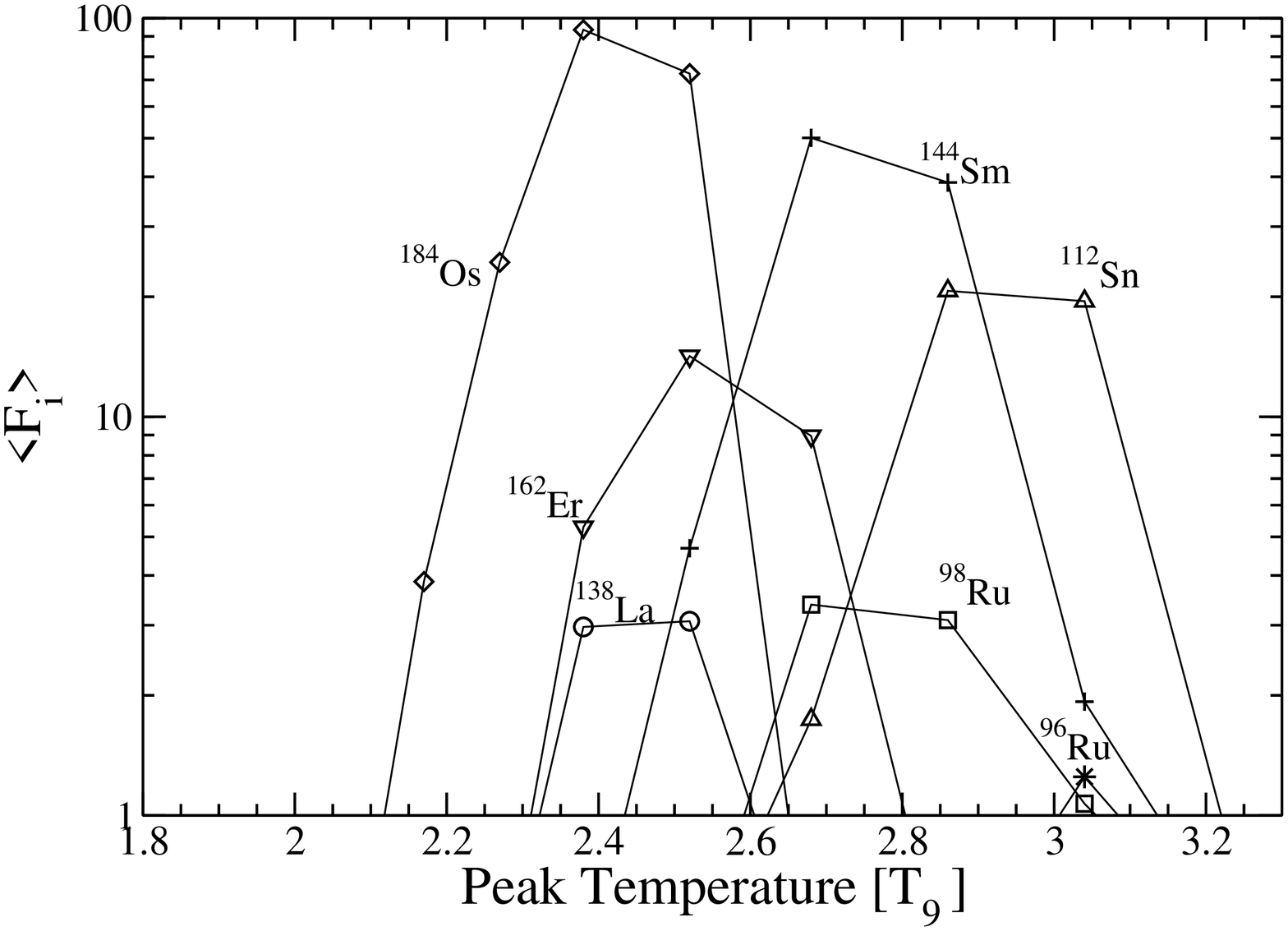}
\caption{Overproduction factors for selected {\it p} nuclei ($^{96}$Ru,$^{98}$Ru,
$^{112}$Sn, $^{138}$La, $^{144}$Sm, $^{162}$Er and $^{184}$Os) as a function
of temperature indicating the production efficiency in the different {\it p}-process
zones.}\label{rapp_fig11}
\end{center}
\end{figure}

The middle part of Fig. \ref{rapp_fig10} displays the sensitivity of the predictions
to global changes in the proton-induced and inverse reaction rates. No impact
can be observed in the mass range $A\ge$110 despite the
fact that the reaction flux pattern shows a pronounced
($\gamma$,p) flux component between $N=$50 and $N=$82. In our model
the only exception is $^{126}$Xe, which is produced by
$\beta$-decay of $^{126}$Ba. The final abundance depends strongly
on the ($\gamma$,p) production chain of $^{126}$Ba in the
higher temperature mass layers. A more pronounced effect is
obtained in the lower mass range $A<$110
where the {\it p} nuclei with $Z\le$50 and $N\ge$50, such as $^{92}$Mo and
$^{96}$Ru, are mostly depleted by ($\gamma$,p) photodisintegration
(Figs. \ref{rapp_fig5} and \ref{rapp_fig9}). A reduction in rate is therefore
directly correlated with an increase in abundance. 
The high neutron and alpha binding energies of the neutron magic nuclei
with $N = 50$ terminate the ($\gamma, n$)-reaction flux at $^{92}$Mo
(see Figs. \ref{rapp_fig6} and \ref{rapp_fig7}). At this point the flux continues only via the
($\gamma, p$) rate because of the relatively low proton binding energy
(see Fig. \ref{rapp_fig8}).
Hence, the $^{92}$Mo abundance is strongly determined by the rate of
the $^{92}$Mo($\gamma, p$) reaction. In contrast, the $^{94}$Mo($\gamma, p$)
reaction  has no effect on the $^{94}$Mo abundance since the much lower neutron
binding energy of this isotope results in a dominance of the ($\gamma, n$)
channel. Similarly, $^{96}$Ru and $^{98}$Ru are both strongly bound against
($\gamma, n$) reactions (see Fig. \ref{rapp_fig8}), but $^{96}$Ru has a sufficiently
small proton binding energy as well to exhibit a clear sensitivity to
changes of the ($\gamma, p$) rate.
Therefore, the final
abundances of the {\it p} nuclei $^{94}$Mo and $^{98}$Ru remain
unaffected by changes in the ($\gamma, p$) rates. 
The abundances of {\it p} nuclei in the low mass range, such
as $^{74}$Se and $^{78}$Kr correlate inversely with the strength
of the proton capture reactions (chapter 4.4).

Global variation of the ($\alpha,\gamma$) and ($\gamma,\alpha$)
rates has a strong impact on the {\it p}-nuclei abundances above
$A=$140 and $N\ge$82 as shown in the bottom panel of Fig.
\ref{rapp_fig10}. In this range the reaction pattern is strongly
affected by ($\gamma,\alpha$) photodisintegration reactions (Figs.
\ref{rapp_fig5} and \ref{rapp_fig9}). With increased rates the
higher mass {\it p} nuclei are bypassed and the material is
processed towards the lower mass {\it p} nuclei such as
$^{144}$Sm, $^{156}$Dy and $^{162}$Er. If the reaction rates are
lower, processing towards these {\it p} nuclei is less efficient
and causes a relative enrichment in the higher mass nuclei such as
$^{174}$Hf, $^{180}$W, and $^{190}$Pt. The abundance of $^{152}$Gd
remains unaffected by changes in the $\alpha$ capture or emission
rates. In the case of $^{164}$Er, variation of the
($\gamma,\alpha$) rates leads always to a decrease in abundance.
If the rate is reduced the direct feeding through
$^{168}$Yb($\gamma,\alpha$)$^{164}$Er is suppressed, and if it is
increased the abundances of higher mass feeding isotopes are
reduced by faster processing towards lower mass $N=$82 isotones.
In the lower mass range the abundance of $^{96}$Ru is by far the 
most sensitive to changes in the ($\gamma,\alpha$) rates. Since the
depletion depends significantly on the $^{96}$Ru($\gamma,\alpha$)
rate, a corresponding enhancement of this rate causes a
substantial reduction in the $^{96}$Ru abundance. This is similar
to the case of enhancing the $^{96}$Ru($\gamma$,p) depletion rate
discussed above.

\begin{figure}
\begin{center}
\includegraphics*[width=9cm]{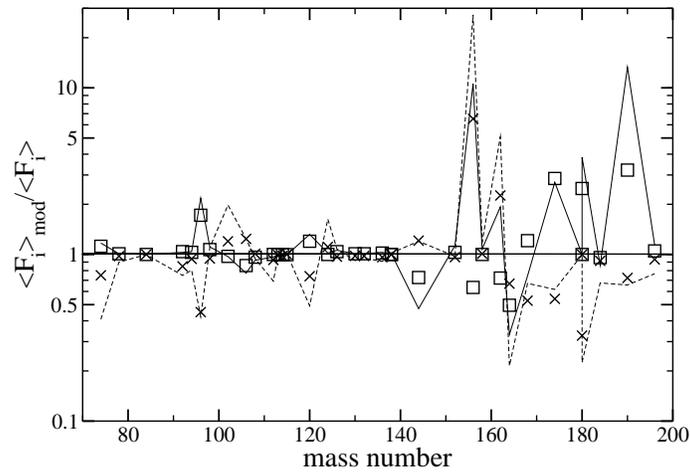}
\caption{Same as bottom panel of Fig. \ref{rapp_fig10}, but with the
($\gamma,\alpha$) rates modified by factors of 0.1 (solid line)
and 10 (dashed line). The results for three times smaller and larger
rates are indicated by squares and crosses, respectively.}
\label{rapp_fig12}
\end{center}
\end{figure}

As pointed out before, experimental studies indicate deviations of
up to one order of magnitude with respect to HF predictions for
$\alpha$ capture and their inverse
photodisintegration processes. We therefore performed a second
{\it p}-process simulation with the ($\gamma,\alpha$) rates modified by
factors of 0.1 and 10, respectively, to account for a broader range
of uncertainty. Fig. \ref{rapp_fig12} shows that these changes result in
a similar pattern as observed in the previous study (bottom panel
of Fig. \ref{rapp_fig10}). A significant change can only be discerned for
$^{156}$Dy, $^{162}$Er, and to a lesser extent for $^{190}$Pt. The
increase with higher reaction rates is due to enhanced feeding of
$^{156}$Dy and $^{162}$Er through ($\gamma$,$\alpha$) reactions in
the mass zones with lower peak temperatures (Fig. \ref{rapp_fig11}).

We have focused so far on global changes of reaction rates.
In the following we discuss the impact of
reactions feeding or depleting only the {\it p} nuclei
in order to determine how critical these
reactions are for the overall abundance predictions.

\subsection{Impact of rates for {\it p}-nuclei reactions}

Similar to the discussion before, the rates directly feeding or
depleting the {\it p} nuclei were changed by factors of three to see if the
calculated {\it p}-process abundances are determined by the global
reaction flux or if they are associated with these specific
reactions. This will have consequences for the
identification of reactions to be selected for further
experimental studies.

Changing the neutron capture rates and the respective
photodissociation rates into the neutron channel by a factor of
three produced only negligible changes in the resulting {\it p}
abundances compared with the results of the global rate changes.
In particular the change of (n,$\gamma$)-rates on the {\it p}
nuclei showed almost no influence while the change of the
($\gamma$,n)-reactions showed mainly a slight sensitivity between
80$<A<$160. Since ($\gamma$,n) photodisintegration processes
provide the main feeding path for the transformation of the
initial seed nuclei towards neutron deficient isotopes no specific
outstanding reactions could be identified in this analysis.
The reason is that the p-process reaction flow proceeds through
neutron deficient nuclei several mass units away from stability.
With increasing neutron separation energy the ($\gamma$,n)
reaction rates decline rapidly and the reaction flow is carried by
($\gamma$,p) and ($\gamma$,$\alpha$) channels. It is important to
identify these branching points which determine the overall
$p$-process reaction flow pattern towards lower masses. The
branching points are clearly temperature dependent and differ
between the different burning zones. The highest sensitivity
appears to be correlated with ($\gamma$,n) reactions near the $N=50$, 82
closed neutron shells where the ($\gamma$,n) reaction flow changes
into a ($\gamma$,p) dominated reaction flow pattern. In addition,
the weak sensitivity of the {\it p} abundances to the
($\gamma$,n) reactions feeding the {\it p} nuclei is
not unexpected. These reactions typically
occur on odd $N$ nuclei with lower Q-values and have therefore
higher rates compared to neighboring ($\gamma$,n) reactions on even
$N$ nuclei. They are therefore not expected to be particularly
important bottle-necks in the {\it p}-process reaction flow.

\begin{figure}
\begin{center}
\includegraphics*[width=9cm]{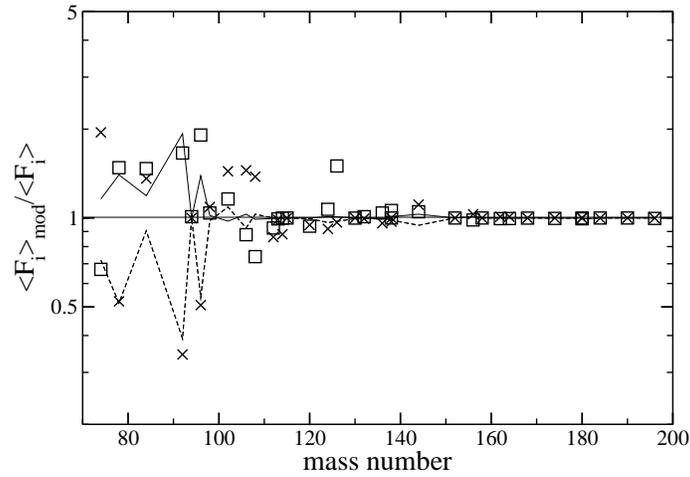}
\caption{Same as middle panel of Fig. \ref{rapp_fig10}, but modifying only the
($\gamma$,p) rates of all {\it p} nuclei and their inverse (p,$\gamma$) rates 
by factors of 1/3
(solid line) and 3 (dashed line). The corresponding results of Fig.
\ref{rapp_fig10} are indicated by squares and crosses, respectively.}
\label{rapp_fig13}
\end{center}
\end{figure}

In the case of ($\gamma$,p) dissociation reactions on
{\it p} nuclei and their inverse processes, 
a modification of these rates has a direct impact on the
abundances of {\it p} nuclei in the mass range $A<$100. Fig. \ref{rapp_fig13} compares
the overproduction factors of {\it p} nuclei based on the modified rates
to the ones based on standard HF rates by
showing the ratio as a function of atomic mass number $A$. The
squares (factor 1/3) and crosses (factor 3) reflect the global changes
of all proton related reaction rates discussed before (see also the mid part
of Fig. \ref{rapp_fig10}), whereas the solid and dashed lines
denote the corresponding ratios obtained by changing only the ($\gamma$,p) 
dissociation rates on {\it p} nuclei and their inverse (p,$\gamma$) reaction rates.
The results indicate that the reactions directly associated with
{\it p} nuclei are the most critical ones,
though in a few cases
($A=74$, 102, 106, 108) reactions on non-$p$ nuclei are also important
(see also table 2).
The figure clearly shows that
the ($\gamma$,p)-rates determine
significantly the {\it p} abundances of $^{74}$Se, $^{78}$Kr, $^{84}$Sr,
$^{92}$Mo, and $^{96}$Ru since the overabundance ratios scale
inversely with the rate scaling factors. Proton capture rates have
almost no influence on the {\it p} abundances, they play only a role for
the light {\it p} nuclei $^{74}$Se and $^{84}$Sr. The
comparison of the overabundance predictions based on the globally
changed rates and selectively changed rates suggests that for
reactions involving protons the individual feeding and depleting
processes of {\it p} nuclei are significant. Fig. \ref{rapp_fig13} indicates
that the abundances of $^{78}$Kr, $^{92}$Mo, and $^{96}$Ru depend
sensitively on the photodisintegration of these nuclei. It has
been pointed out before that particularly the
$^{92}$Mo($\gamma$,p)$^{91}$Nb rate determines the $^{92}$Mo/$^{92}$Nb
abundance ratio predicted by supernova models \citep{DRM03}.

Also in the case of $\alpha$ capture and $\alpha$ disintegration
reactions the results indicate that the produced overabundances
are closely correlated to the individual feeding and depleting
reactions of the light {\it p} nuclei. 
Correlations in
specific cases have already been found in Fig. \ref{rapp_fig12}
with ($\gamma$,$\alpha$) reactions depleting $^{74}$Se, $^{96}$Ru,
$^{120}$Te, $^{122}$Xe and ($\gamma$,$\alpha$) reactions feeding
$^{102}$Pd, $^{106}$Cd, $^{144}$Sm, $^{156}$Dy and $^{162}$Er as
discussed in the chapter 4.1.

The present simulations yield low abundances for the Mo and Ru
{\it p} isotopes, in agreement with previous calculations. 
The difficulty in solving this problem within the 
astrophysical model and seed distribution discussed here
comes from the fact that $^{92,94}$Mo and $^{96,98}$Ru
make up about 43\% of the total solar $p$ nucleus
abundances. 
Therefore, increasing their overproduction significantly 
requires a very efficient conversion of a significant
fraction of all the heavier seed nuclei into Mo and Ru. As Fig.~\ref{rapp_fig11}
shows, an absolute overproduction factor of the order of $10-90$
would be needed for the Ru and Mo $p$ isotopes to be in line 
with heavier $p$ nuclei, assuming that specific isotopes are
roughly produced in
layers of similar mass. 
For comparison, conversion of the entire $Z>42$ seed into 
$^{92,94}$Mo and $^{96,98}$Ru would be required to 
bring their overproduction factors up to an average of 
70. Even if $^{92,94}$Mo would be produced in separate
layers without coproduction of $^{96,98}$Ru the maximum 
achievable overproduction factor would only be around 110. 
In contrast to the earlier findings of \citet{RPA90} using 
a parametrized $p-$process model 
this would be in principle sufficient, but it is difficult to see how 
such a major increase in Mo and Ru production efficiency could come about.
Nevertheless, to investigate this issue further, 
we decreased the $^{92,94}$Mo($\gamma $,p) and
$^{96,98}$Ru($\gamma $,p) rates to 10\% of their HF prediction.
These reaction rates are the main destruction mechanism for 
$^{92,94}$Mo and $^{96,98}$Ru.
The resulting change in abundance is shown in
Fig.~\ref{rapp_fig15}. Only a moderate enhancement of the associated
abundances is observed. 
Accordingly, reaction rates are not
responsible for the underproduction of the light {\it p} nuclei.

\subsection{Critical {\it p}-process reaction rates}

The simulations discussed in the previous sections indicate that
within the present model relatively few selected reaction rates
have a major impact on the final {\it p} abundances. The critical
rates are typically associated with a strong feeding flux or with
a particular branching in the reaction flux feeding or bypassing
the {\it p} nuclei (Fig. \ref{rapp_fig5}). The ($\gamma$,n)
reactions control the overall feeding of the {\it p}-process flux
from the weak {\it s}-process seed distribution, thus affecting
the final {\it p}-process abundance distribution over the entire
mass range. This influence is particularly visible in the range
above $N=$82 as suggested by Fig. \ref{rapp_fig10} (top panel). In
the following we want to concentrate on the discussion of the
($\gamma$,p), (n,p), and ($\gamma,\alpha$) branchings which divert
the {\it p}-process flux towards lower masses. We identified a set
of ($\gamma$,p) and (n,p) reactions and the respective inverse
processes to investigate their specific impact on the simulations.
Table \ref{tab2} also includes those reactions in the mass range
70$<A<$78, which seem to affect the abundances of the very light
{\it p} nuclei below $^{92}$Mo.

\begin{figure}
\begin{center}
\includegraphics*[width=9cm]{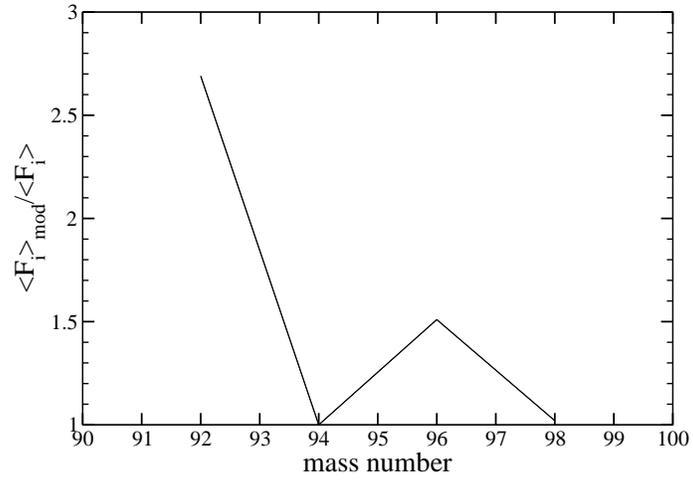}
\caption{
Ratio of the overproduction factors for the
{\it p} abundances in the mass region 92$\leq A \leq 98$
obtained when reducing the ($\gamma$,p) destruction rates of
$^{92}$Mo, $^{94}$Mo $^{96}$Ru, and $^{98}$Ru by a factor of ten
to the overproduction factors obtained with unchanged rates.}
\label{rapp_fig15}
\end{center}
\end{figure}

A change in the rates by a constant factor can result in opposite
sensitivities as can be seen for $^{74}$Se and $^{78}$Kr in the middle
panel of Fig. \ref{rapp_fig10}. For proton induced reactions and their inverse
reactions the $^{78}$Kr abundance depends only on the depletion
reaction $^{78}$Kr($\gamma, p$) (see top part of Fig. \ref{rapp_fig15}),
hence an increase of
the proton-induced rates results always in a lower $^{78}$Kr abundance.
In contrast, the $^{74}$Se abundance depends not only on the
$^{74}$Se($\gamma, p$) rate since it is produced in the high temperature
layers where the flow via $^{74}$Ge($p, \gamma$)$^{75}$As and
$^{75}$As($p, n$)$^{75}$Se($\gamma, n$) leads to an enhancement of the $^{74}$Se
abundance as these rates are increased.

\begin{figure}
\begin{center}
\includegraphics*[width=9cm]{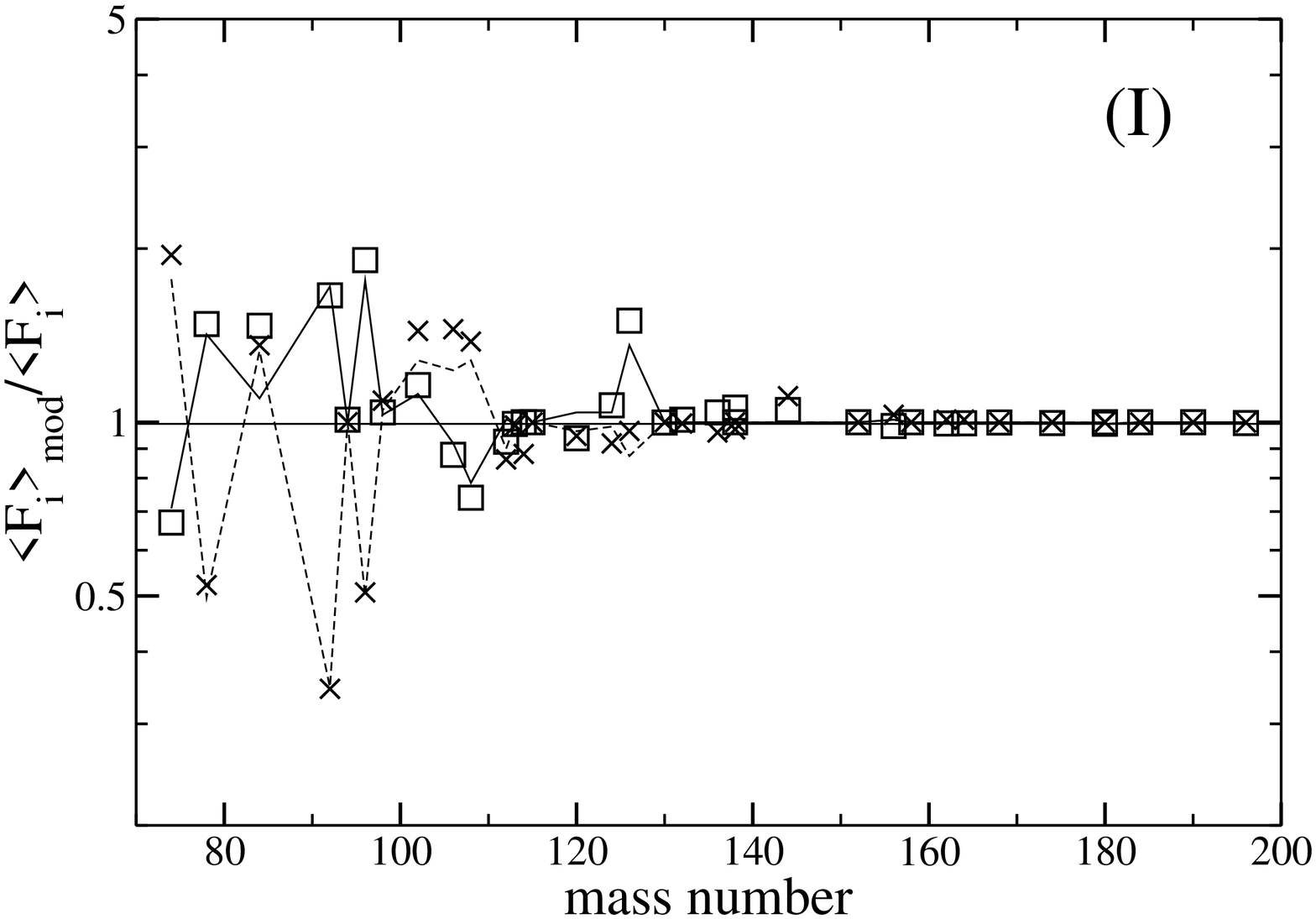}
\includegraphics*[width=9cm]{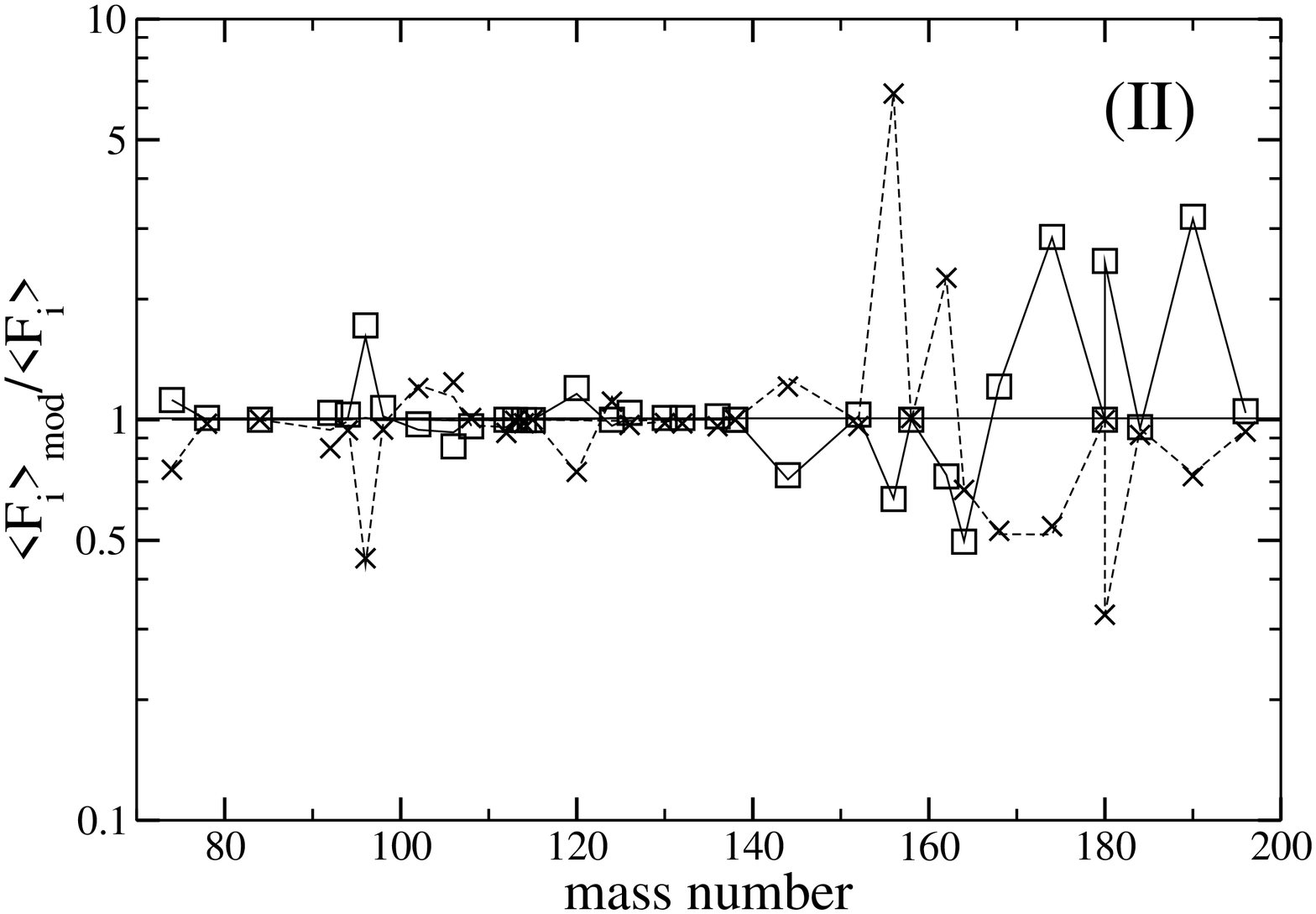}
\caption{Influence of the selected reactions listed in Tables \ref{tab2}
and \ref{tab3} compared to the results obtained by modification of all
accepted HF-rates as shown in Fig. \ref{rapp_fig10} (symbols). The top panel
refers to the proton reactions of Table \ref{tab2}, and the bottom panel to the
$\alpha$ reactions of Table \ref{tab3}. The effect of changes in the
selected reactions is indicated by 
solid and dashed
lines for reducing and
increasing the rates by a factor of three, respectively. It is obvious
that the selected rates are almost completely accounting for the response
of the $p$-process network to reaction rate uncertainties.}
\label{rapp_fig16}
\end{center}
\end{figure}

The influence of the selected ($\gamma$,p) and (p,n) reactions as
well as their inverse processes is demonstrated in the top panel
of Fig. \ref{rapp_fig16}, where the abundance modifications
resulting from the global change of all ($\gamma$,p) and (n,p)
reaction rates and their inverse processes
by a factor of three (mid panel of Fig.
\ref{rapp_fig10}) are compared with the ones based on the change
of the rates listed in Table \ref{tab2}. The solid and dotted
lines indicate the consequences of increasing and decreasing these
rates by a factor of three, respectively. Within the general
uncertainties the observed abundance pattern agrees very well with
the abundance pattern resulting from a global change of all rates
within the same boundaries and shows that the impact of
($\gamma$,p) photodisintegration processes is confined to the {\it
p}-process flux in the range $Z<$50 with the only exception of
$^{126}$Ba($\gamma$,p)$^{125}$Cs. The impact of the ($\gamma$,p)
reactions on the abundance predictions for the light {\it p}
nuclei $^{92}$Mo and $^{96}$Ru is substantial but - within the
chosen boundaries - not strong enough to explain the large
under-production of these crucial {\it p} nuclei. Sensitivities to 
the uncertainties in
the reaction rates of the (p,n) processes are mainly observed in
the {\it p}-process range $N \le$50 where photodisintegration plays
a less dominant role than in the higher mass regions.

The high seed abundances in the mass region from Hg to Pb (Fig.
\ref{rapp_fig3})
are efficiently shifted by ($\gamma, \alpha$) reactions towards lower
masses, particularly at temperatures of $T_9 = 2.6$. This mass flow
is considerably enhanced by increasing the rates of the relevant
reactions listed in table 3. For example, the abundances of
the $p$ nuclei $^{156}$Dy and $^{162}$Er exhibit a strong sensitivity
on the $^{160}$Er($\gamma, \alpha$)$^{156}$Dy and $^{166}$Yb($\gamma,
\alpha$)$^{162}$Er reactions, respectively. As shown in Fig. \ref{rapp_fig15},
increasing and decreasing these rates by factors of three results
in a corresponding decrease and increase of the overproduction
factors of $^{156}$Dy and $^{162}$Er. However, the global change
of the ($\gamma, \alpha$) rates by a factor of ten as illustrated
in Fig. \ref{rapp_fig12} leads to an inverse trend in the overproduction factors
for $^{156}$Dy and $^{162}$Er, which are strongly enhanced also if the
rates are increased. This behavior reflects the enhanced mass flow
from the abundant seeds in the Hg to Pb region. In summary, the
importance of the ($\gamma, \alpha$) reactions calls for more
systematic studies in the mass region $A \geq 140$.

Table \ref{tab3} lists the ($\gamma,\alpha$) reactions, which carry
a substantial reaction flux in the {\it p} process (see Fig. \ref{rapp_fig5}).
Again, modifying these reactions within the given uncertainty
range causes significant changes in the abundances of the {\it p} nuclei.
This is demonstrated in the lower panel of Fig. \ref{rapp_fig16}
by comparing the abundance modifications resulting from the global
change of all ($\gamma,\alpha$) reactions in the network by a
factor of three (bottom panel of Fig. \ref{rapp_fig10}) with the ones
based only on the modified rates listed in Table \ref{tab3}.
Again, the solid and dotted lines indicate the response to an increase
and the reduction of these rates by a factor of three, respectively.

The changes in {\it p}-nuclei abundances based on the modification
of the selected rates in Table \ref{tab3} agree well with the
global change of all ($\gamma,\alpha$) rates. The figure supports
the argument that the impact of the ($\gamma,\alpha$)
photodisintegration processes is mainly visible in the higher mass
{\it p}-process range above $N=$82. This underlines the particular
importance of ($\gamma,\alpha$) rates for the mass range $N \ge$82
except for the case of the light {\it p} 
nuclei $^{96}$Ru and $^{74}$Se, where
the predicted abundance is correlated with the depletion via
$^{96}$Ru($\gamma,\alpha$)$^{92}$Mo and $^{74}$Se($\gamma$,$\alpha$)$^{70}$Ge
respectively, as 
indicated in Fig. \ref{rapp_fig5}.

The analysis of the abundances for {\it p} nuclei with $N \ge$82
indicates that the low abundances of the {\it p} nuclei
$^{156}$Dy, $^{162}$Er, and $^{190}$Pt might be due to the
($\gamma,\alpha)$ rates in that mass range, while {\it p} nuclei
such as $^{152}$Gd, $^{158}$Dy, and $^{164}$Er depend more on the
strength of the particular ($\gamma$,n) feeding from the seed
abundance distribution.

The results shown in Fig. \ref{rapp_fig16} suggest that comparably
few rates involving charged particles have a direct impact on the
abundances of the {\it p} nuclei. These critical rates, which are
summarized in Tables \ref{tab2} and \ref{tab3}, carry most of the charged particle
related uncertainties for simulations within the discussed {\it
p}-process model frame. Since all of these rates are presently
based on global HF calculations \citep{RaT00,RaT01,RaT04},
experimental confirmation is necessary for reducing the inherent
nuclear model uncertainties. This serves not only the purpose of
obtaining better data, but since {\it p}-process abundances are
sensitive to changes in these specific rates, a better knowledge
of those rates will help to improve the fine-tuning for modeling
of the supernova shock front traversing the O-Ne shell of the
pre-supernova star.

Our list of potentially important reactions largely differs from the one given
by \citet{Rau06}. This is not surprising because the latter work
identifies reaction flow branchings in a more simplified approach
and does not follow the hydrodynamic evolution of different
p-process layers. In addition, we have not limited
our analysis to branchings,
and we have also included other types of reactions such as (n,p) or (p,n).

\begin{figure}
\begin{center}
\includegraphics*[width=9cm]{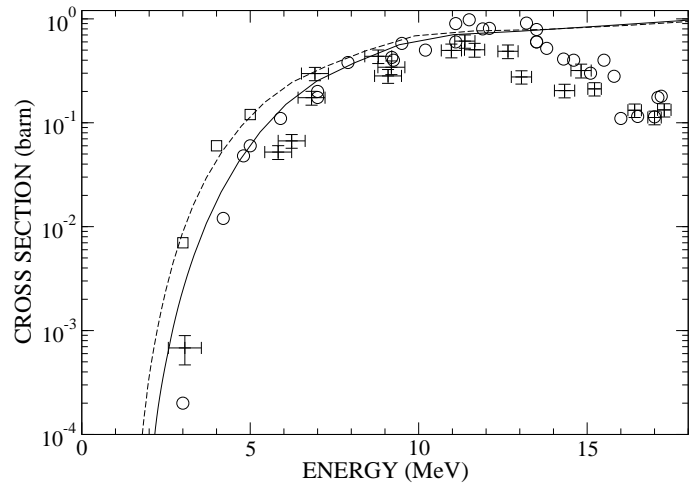}
\caption{Cross section data for two sets of $^{75}$As(p,n)
measurements ((squares) \citep{KMG79}, (circles) \citep{MQS88}) in
comparison with Hauser Feshbach predictions (solid line) \citep{RaT01}. Also
shown is the experimental cross section of $^{85}$Rb(p,n)
((crosses) \citep{KQN02}) in comparison with Hauser Feshbach
predictions (dashed line) \citep{RaT01}.\label{rapp fig17}}
\end{center}
\end{figure}

There are experimental data for some of the listed (p,n) reactions.
Fig. \ref{rapp fig17} shows the experimental data for the
$^{75}$As(p,n) and the $^{85}$Rb(p,n) reaction in comparison with
Hauser Feshbach predictions. Only three data points have been
determined for $^{75}$As(p,n) in the energy range from 3 to 5 MeV
\citep{KMG79}. The experimental results are in excellent agreement
with the statistical Hauser Feshbach predictions \citep{RaT01}. On
the other hand, the $^{75}$As(p,n) cross sections obtained by a
measurement of $^{75}$As(p,xn) reactions in the energy range of 3
- 45 MeV \citep{MQS88} show significant discrepancies to the model
predictions for the astrophysical low energy range. Experimental
cross section data for the $^{85}$Rb(p,n) reaction are available
for a wide energy range from 3 MeV up to 100 MeV \citep{KQN02}.
The experimental results are on average lower than the  Hauser
Feshbach predictions \citep{RaT01}; on the other side, the data
are handicapped by huge experimental uncertainties. In both cases
independent experimental verification of the cross sections for
the astrophysical energy range is desirable.

Different experimental techniques will be necessary for a complete
study of these critical reactions. As mentioned before, in
previous work the experimental effort concentrated on the study of
capture reactions using the activation technique and in a few
cases on direct photodisintegration studies with photon beams. The
activation technique for capture reactions on stable nuclei is
limited to cases where the final reaction products are
characterized by specific $\gamma$-radiation signatures. This is
necessary for uniquely identifying the reaction products and
separating them from the large background activity originating
from target impurities. This technique for example can be applied
for the measurement of $^{75}$As(p,n)$^{75}$Se,
$^{106}$Cd($\alpha,\gamma$)$^{110}$Sn, or
$^{196}$Hg($\alpha,\gamma$)$^{200}$Pb, as well as a larger set of
(n,$\gamma$) cross sections.

If the characteristic activity is masked by background or if the
reaction product lacks characteristic activity, alternative
detection methods can be envisioned through AMS methods where the
reaction products are chemically separated from the activated
target and analyzed through high resolution accelerator mass
spectroscopy. This approach has been successfully applied in the
study of the $^{62}$Ni(n,$\gamma$)$^{63}$Ni $s$-process reaction
\citep{NPA04}. Possible examples for applying this specific
technique are the reactions $^{72}$Ge(p,$\gamma$)$^{73}$As, or for
the higher mass range $^{156}$Dy($\alpha,\gamma$)$^{160}$Er or
$^{190}$Pt($\alpha,\gamma$)$^{194}$Hg, which are characterized by
such a low Q$_{\beta}$ value that a measurement of the decay
activity is difficult.

Inverse photodisintegration measurements have also been
developed as a powerful experimental tool
\citep{UYA01,VMB01}. 
The increasing availability of
high energy photon beams at the HIGS facility at TUNL
(http://higs.tunl.duke.edu/) or at the ELBE facility of the
Forschungszentrum Rossendorf\\
(http://www.fz-rossendorf.de/pls/rois/Cms?pNid=144) opens new
opportunities for targeting reactions such as
$^{74}$Se($\gamma$,p)$^{73}$As, $^{92}$Mo($\gamma$,p)$^{91}$Nb,
and $^{96}$Ru($\gamma$,p)$^{95}$Tc, which directly affect the
light {\it p}-nuclei abundances. Again, the reaction product can
be detected by measuring the characteristic activity. If the
latter is masked by background or falls below present detection
limits alternative analytical methods such as the AMS approach can
be applied.

Many of the reactions listed in Tables \ref{tab2} and \ref{tab3}, however, are
photodisintegration processes on radioactive nuclei leading to
radioactive nuclei; reactions with significant impact on flux and
abundances are $^{110}$Sn($\gamma$,p)$^{109}$In,
$^{126}$Ba($\gamma$,p)$^{125}$Cs, or in the range $N \ge$82,
for example $^{152}$Dy($\gamma,\alpha$)$^{148}$Gd,
$^{160}$Yb($\gamma,\alpha$)$^{156}$Er. Such reactions can be
approached by Coulomb dissociation techniques with radioactive
beams. In the specific {\it p}-process cases listed here heavy ion
radioactive beams have to be developed in the energy range of 2-12
MeV/amu depending on the associated Q-values. Coulomb dissociation
techniques are typically applied for studies with light
radioactive particles, which are produced through beam
fragmentation reactions. Such cases are suited for facilities such
as the NSCL at MSU, at RIKEN, and at GSI where heavy neutron
deficient radioactive particles can be produced by spallation or
heavy ion evaporation reactions at inverse kinematics.

\section{SUMMARY}

We have investigated the reaction flux patterns of the {\it p}
process within the framework of a multi-mass zone type II
supernova shock front model. Particular attention was paid to the
influence of {\it p}-process reaction rates within the uncertainty
limits of the theoretical Hauser-Feshbach model. The predicted
{\it p} abundances are similar to the ones observed by previous
work in this field. It could be shown that the endemic
underproduction of the light {\it p} nuclei $^{92,94}$Mo and
$^{96,98}$Ru are not due to the uncertainties in the thermonuclear
reaction rates and must be traced back to other reasons. 
Possible neutron poison reactions in core helium burning and
secondary neutron sources during core carbon burning could modify
significantly the weak {\it s}-process seed abundance distribution
for the {\it p}-process. This was not considered in the present
paper. It should be addressed in future studies to investigate
systematically the {\it s}-process related uncertainties for {\it
p}-process simulations.

Since ($\gamma$,n) reactions control the overall feeding of the
{\it p}-process flux, the corresponding reaction rates affect the
final {\it p}-process abundance distribution over the entire mass
range. This influence is particularly visible in the range above
$N=$82. On the other hand we found that the impact of ($\gamma$,p)
and ($\gamma,\alpha$) reactions is limited to specific mass
regions. Changes in the ($\gamma$,p) reactions translate to direct
modifications of the {\it p}-process abundances in the lower mass
range with $Z\le$50, but similar correlations were not observed in
the higher mass range. In contrast, variations in the
($\gamma,\alpha$) reaction rates impact the abundance predictions
in the higher mass range above $N=$82, while only small effects can
be observed at lower masses. The overall impact of reaction rates
seems not as dramatic as for other processes, the here discussed
modifications of the rates translated in general to a change in
the absolute abundances of a factor two to three. 
This will
directly affect the overproduction factors displayed in Fig.~\ref{rapp_fig4}. 
Improved data are necessary since a reduction of
the associated uncertainty will help to identify more clearly
other uncertainty factors in the present simulation. This includes
the uncertainties resulting from the initial seed distribution or
uncertainties yielding from insufficient model descriptions of the
p-process scenario.

Past experiments have confirmed that theoretically predicted {\it
p}-process reaction rates agree within a factor of two with
experimentally determined proton and neutron capture and the
respective photodisintegration data. For the corresponding $\alpha$
rates considerably larger differences of up to a factor of ten
were reported. More experimental work is necessary to test the
validity of the $\alpha$ potentials used in the model calculations.
It is also necessary to expand the experimental work towards {\it
p}-process reaction studies on neutron deficient nuclei to verify
the reliability of the theoretical predictions for reactions
involving radioactive nuclei. 

\section{ACKNOWLEDGEMENT}

We would like to thank T. Yoshida from the Astronomical Data
Analysis Center, National Astronomical Observatory in Osawa,
Mitaka, Tokyo, (Japan) for providing us with the temperature and
density profiles and M. Rayet from Free University of Brussels
(Belgium) for sending us the $s$-process seed distribution. 
We also thank F.-K. Thielemann for providing the reaction 
network solver.
This
project is supported through the Joint Institute of Nuclear
Astrophysics by NSF-PFC grant PHY02-16783. 
H.S. acknowledges 
additional support through NSF grant PHY01-10253.


\clearpage
\begin{table}
\caption{Mass coordinate, maximum temperature and maximum density in the
investigated $p$-process layers.\label{tab1}}
\begin{center}
\begin{tabular}{cccc}
\tableline\tableline
layer & mass inside the shell & maximum temperature & maximum density\\
No.   & (M$_{\odot}$) &(T$_{9}$) & 10$^{5}$ (g/cm$^{3})$\\
\tableline
 1& 1.9336& 3.45& 7.85 \\
 2& 1.9658& 3.11& 6.64 \\
 3& 2.0085& 2.96& 5.68 \\
 4& 2.0508& 2.76& 4.82 \\
 5& 2.1037& 2.60& 4.07 \\
 6& 2.1564& 2.44& 3.56 \\
 7& 2.2090& 2.32& 3.16 \\
 8& 2.2614& 2.21& 2.80 \\
 9& 2.3136& 2.12& 2.54 \\
 10& 2.3655& 2.04& 2.30 \\
 11& 2.4171& 1.97& 2.11 \\
 12& 2.4684& 1.91& 1.94 \\
 13& 2.5249& 1.84& 1.75\\
 14& 2.5825& 1.79& 1.68\\
\tableline
\end{tabular}
\end{center}
\end{table}

\clearpage

\begin{table}
\caption{
Selected ($\gamma$,p) or (n,p) reactions
that together with their respective inverse reactions were found to
exhibit the strongest influence on the final $p$ abundances. Their
impact is illustrated in the top panel of Fig. \ref{rapp_fig16}. Particularly 
important rates are marked with $^{*)}$ .\label{tab2}}
\begin{center}

\begin{tabular}{lclclccc}
\tableline
$^{126}$Ba($\gamma$,p)$^{125}$Cs$^{*)}$ & &  $^{92}$Mo($\gamma$,p)$^{91}$Nb$^{*)}$  & & $^{75}$Se(n,p)$^{75}$As$^{*)}$  \\
$^{110}$Sn($\gamma$,p)$^{109}$In$^{*)}$ & &  $^{86}$Rb(n,p)$^{86}$Kr$^{*)}$         & & $^{74}$Se($\gamma$,p)$^{73}$As$^{*)}$  \\
$^{106}$Cd($\gamma$,p)$^{105}$Ag        & &  $^{85}$Sr(n,p)$^{85}$Rb$^{*)}$         & & $^{76}$As(n,p)$^{76}$Ge$^{*)}$  \\
$^{104}$Cd($\gamma$,p)$^{103}$Ag        & &  $^{84}$Sr($\gamma$,p)$^{83}$Rb$^{*)}$  & & $^{75}$As($\gamma$,p)$^{74}$Ge \\
$^{100}$Pd($\gamma$,p)$^{99}$Rh         & &  $^{78}$Kr($\gamma$,p)$^{77}$Br$^{*)}$  & & $^{73}$As($\gamma$,p)$^{72}$Ge$^{*)}$  \\
$^{96}$Ru($\gamma$,p)$^{95}$Tc$^{*)}$   & &  $^{77}$Se(n,p)$^{77}$As                & & $^{71}$Ge(n,p)$^{71}$Ga \\
\tableline
\end{tabular}
\end{center}
\end{table}

\clearpage

\begin{table}
\caption{
Selected 
($\gamma,\alpha$) 
reaction chains,
which were found to exhibit the strongest influence on the final $p$
abundances. Their impact is illustrated in the bottom panel of Fig. \ref{rapp_fig16}.
Particularly important rates are marked with $^{*)}$. \label{tab3}}
\begin{center}
\begin{tabular}{lcl}
\tableline
  $^{74}$Se($\gamma$,$\alpha$)$^{70}$Ge$^{*)}$    &  &  $^{196}$Pb($\gamma$,$\alpha$)$^{192}$Hg$^{*)}$  \\
  $^{96}$Ru($\gamma$,$\alpha$)$^{92}$Mo$^{*)}$    &  &  $^{191}$Hg(2$\gamma$,2$\alpha$)$^{183}$Os \\
  $^{106}$Cd($\gamma$,$\alpha$)$^{102}$Pd         &  &  $^{190}$Hg(3$\gamma$,3$\alpha$)$^{178}$W$^{*)}$   \\
  $^{110}$Sn($\gamma$,$\alpha$)$^{106}$Cd$^{*)}$  &  &  $^{189}$Hg(3$\gamma$,3$\alpha$)$^{177}$W  \\
  $^{120}$Te($\gamma$,$\alpha$)$^{116}$Sn$^{*)}$  &  &  $^{188}$Hg(5$\gamma$,5$\alpha$)$^{168}$Yb$^{*)}$  \\
  $^{122}$Xe($\gamma$,$\alpha$)$^{118}$Te         &  &  $^{183}$Pt(3$\gamma$,3$\alpha$)$^{171}$Hf \\
  $^{128}$Ba($\gamma$,$\alpha$)$^{124}$Xe$^{*)}$  &  &  $^{178}$Os(4$\gamma$,4$\alpha$)$^{162}$Er$^{*)}$  \\
                                          &  &  $^{177}$Os(3$\gamma$,3$\alpha$)$^{165}$Yb \\
                                          &  &  $^{176}$Os(5$\gamma$,5$\alpha$)$^{156}$Dy$^{*)}$  \\
                                          &  &  $^{167}$Hf($\gamma$,$\alpha$)$^{163}$Yb \\
                                          &  &  $^{166}$Hf(5$\gamma$,5$\alpha$)$^{146}$Sm$^{*)}$  \\
                                          &  &  $^{156}$Er(3$\gamma$,3$\alpha$)$^{144}$Sm$^{*)}$  \\
\tableline
\end{tabular}
\end{center}
\end{table}

\clearpage

\clearpage

\clearpage

\clearpage

\clearpage

\clearpage

\clearpage

\clearpage

\clearpage

\clearpage


\clearpage

\clearpage

\clearpage

\end{document}